\documentclass{aa}
\usepackage[varg]{txfonts}

\titlerunning{Microwave Thermal Emission from the Zodiacal Dust Cloud}
\authorrunning{V. V. Dikarev \& D. J. Schwarz}

\begin{document}


\title{The Microwave Thermal Emission\\from the Zodiacal Dust Cloud\\
       Predicted with Contemporary Meteoroid Models}


\author{Valery V. Dikarev \and Dominik J. Schwarz}
\institute{Faculty of Physics, Bielefeld University, Postfach 100131, 33501 Bielefeld, Germany\\
\email{vdikarev@physik.uni-bielefeld.de}}

\date{Received <date> / Accepted <date>}



\abstract{Predictions of the microwave thermal emission from the interplanetary dust cloud are
made using several contemporary meteoroid models to construct the distributions
of cross-section area of dust in space, and applying the Mie light-scattering theory to estimate
the temperatures and emissivities of dust particles in broad size and heliocentric distance ranges.
In particular, the model of the interplanetary dust cloud by Kelsall et al. (1998, ApJ~508:~44--73),
the five populations of interplanetary meteoroids of Divine (1993, JGR~98(E9): 17,029--17,048)
and the Interplanetary Meteoroid Engineering Model (IMEM) by Dikarev et al. (2004, EMP~95: 109--122)
are used in combination with the optical properties of olivine, carbonaceous and iron spherical particles.
The Kelsall model has been widely accepted by the Cosmic Microwave Background (CMB) community.
We show, however, that it predicts the microwave emission from interplanetary dust
remarkably different from the results of application of the meteoroid engineering models.
We make maps and spectra of the microwave emission predicted by the three models assuming
variant composition of dust particles. Predictions can be used to look for the emission
from interplanetary dust in CMB experiments as well as to plan new observations.}


\maketitle


\section{Introduction}

The unprecedented high-precision surveys of the microwave sky
performed recently by the {\it Wilkinson Microwave Anisotropy Probe} \citep[WMAP,][]{Bennett-et-al-2013ApJS}
and {\it Planck} \citep{Planck-Results-I-2014} observatories
in search and characterisation of the large- and small-scale structure
of the Cosmic Microwave Background (CMB)
anisotropies pose new challenges for simulation and subtraction of the foreground
emission sources, including the Solar-system dust. Previous templates
designed for this purpose were based on the \citet{Kelsall-et-al-1998} model.
They indicated little significance of the interplanetary dust for the {\it WMAP}
survey \citep{Schlegel-et-al-1998}, yet remarkable contribution was
detected at high frequencies of {\it Planck}~\citep{KenGanga-2013}.


The angular power spectrum of CMB anisotropies is in good agreement with the
inflationary $\Lambda$-cold dark matter model \citep{Planck-Results-I-2014}.
However, the reconstructed CMB maps at the largest angular scales reveal 
some intriguing anomalies, among them unexpected alignments of multipole moments, 
in particular with directions singled out by the Solar system~\citep{Schwarz-et-al-2004PhRvL}.
The quadrupole and octopole are found to be mutually aligned
and they define axes that are unusually perpendicular to the ecliptic pole and
parallel to the direction of our motion with respect to the rest frame of the CMB (the
dipole direction). For reviews see
\citet{Bennett-et-al-2011ApJS},
\cite{Copi-et-al-2010} and 
\cite{Bennett-et-al-2013ApJS} and
\citet{Planck-Results-XXIII-2014} as well as
\citet{2013arXiv1311.4562C}
for a detailed analysis of final WMAP and first Planck data. 
It has been suggested that an unaccounted observation bias may persist, e.g.\ yet
another foreground source of the microwave emission bound to the Solar system.
\citet{Dikarev-et-al-2009} explored the possibility that the Solar-system
dust emits more indeed than it was previously thought, and found that
the macroscopic ($>0.1$~mm in size) meteoroids can well contribute $\sim10\;\mu$K
in the microwaves, i.e.\ comparable with the power of the CMB anomalies,
without being detected in the infrared (IR) light and included in the IR-based
models like that of \citet{Kelsall-et-al-1998}.
\citet{Babich-et-al-2007ApJ} and \citet{Hansen-et-al-2012} have also studied
possible contribution of the Kuiper belt objects to the microwave foreground
radiation. \citet{Dikarev-et-al-2008} constructed and tested against the {\it WMAP}
data some dust emission templates.

Here we improve and extend preliminary estimates of \citet{Dikarev-et-al-2009}.
In addition to the Kelsall model, we use two meteoroid engineering models
to make accurate and thorough maps of the thermal emission from
the Zodiacal cloud in the wavelength range from 30~$\mu$m to 30~mm.

The paper is organized as follows. Section~\ref{Meteoroid Models} introduces
three models of the interplanetary meteoroid environment that we use to predict
the thermal emission from the zodiacal dust cloud. A detailed description of
the theory and observations incorporated in each model is provided, it may be helpful
not only for cosmologists interested in understanding the solar-system
microwave foregrounds, but also for developers of new meteoroid models
willing to comprehend earlier designs.
Section~\ref{Thermal Emission Models} describes the thermal
emission models. Empirical models as well as the Mie light-scattering theory
are used to calculate the temperatures and emission intensities
of spherical dust particles composed of silicate, carbonaceous and iron materials,
for broad ranges of size and heliocentric distance.
The models of the spatial distribution of dust and thermal emission
are combined in Section~\ref{The Microwave Thermal Emission from Dust}
in order to make all-sky maps and spectra of the thermal emission
from the interplanetary dust. Conclusions are made in Sect.~\ref{Conclusion}.

Throughout this paper, we will be dealing with the wavelengths mostly,
since the dust optical properties are naturally described in terms of linear scales.
The CMB community, however, is more accustomed to frequencies. A conversion table
is therefore useful for the observation wavebands of infrared detectors and radiometers (Table~\ref{Convert}).
\begin{table}[t]
\caption{Frequencies~$\nu$, wavelengths~$\lambda$ and bandwidths~$\Delta\nu/\nu$
of the CIB/CMB observations with {\it COBE} DIRBE wavelength bands~4 through~10,
{\it WMAP}, and {\it Planck}.\label{Convert}}
\medskip
\begin{tabular}{rrrl}
\hline
Instrument & \multicolumn{1}{c}{$\nu$} & \multicolumn{1}{c}{$\lambda$} & $\Delta\nu/\nu$ \\
\hline
{\it COBE} 
& 61.2~THz  & 4.9~$\mu$m & 0.13 \\
{DIRBE}    
& 25.0~THz  &  12~$\mu$m & 0.54 \\
& 12.0~THz  &  25~$\mu$m & 0.34 \\
& 5.00~THz  &  60~$\mu$m & 0.46 \\
& 3.00~THz  & 100~$\mu$m & 0.32 \\
& 2.14~THz  & 140~$\mu$m & 0.28 \\
& 1.25~THz  & 240~$\mu$m & 0.40 \\
\hline
{\it Planck}      & 857~GHz   & 350~$\mu$m & 0.25 \\
{HFI}             & 545~GHz   & 550~$\mu$m & 0.25 \\
            & 353~GHz   & 850~$\mu$m & 0.25 \\
            & 217~GHz   & 1.4~mm & 0.25     \\
            & 143~GHz   & 2.1~mm & 0.25     \\
            & 100~GHz   & 3.0~mm & 0.25     \\
\hline
{\it Planck}      & 70~GHz & 4.3~mm  & 0.2   \\
{LFI}       & 44~GHz & 6.8~mm  & 0.2     \\
            & 30~GHz & 10.0~mm  & 0.2    \\
\hline
{\it WMAP}~W\phantom{a}    & 94~GHz & 3.2~mm & 0.22    \\
     V\phantom{a}    & 61~GHz & 4.9~mm & 0.23    \\
     Q\phantom{a}    & 41~GHz & 7.3~mm & 0.20    \\
     Ka              & 33~GHz & 9.1~mm & 0.21    \\
     K\phantom{a}    & 22~GHz & 13.6~mm & 0.24   \\
\hline
\end{tabular}
\end{table}

\section{Meteoroid Models}\label{Meteoroid Models}

\begin{figure*}
\centerline{\includegraphics[width=\textwidth]{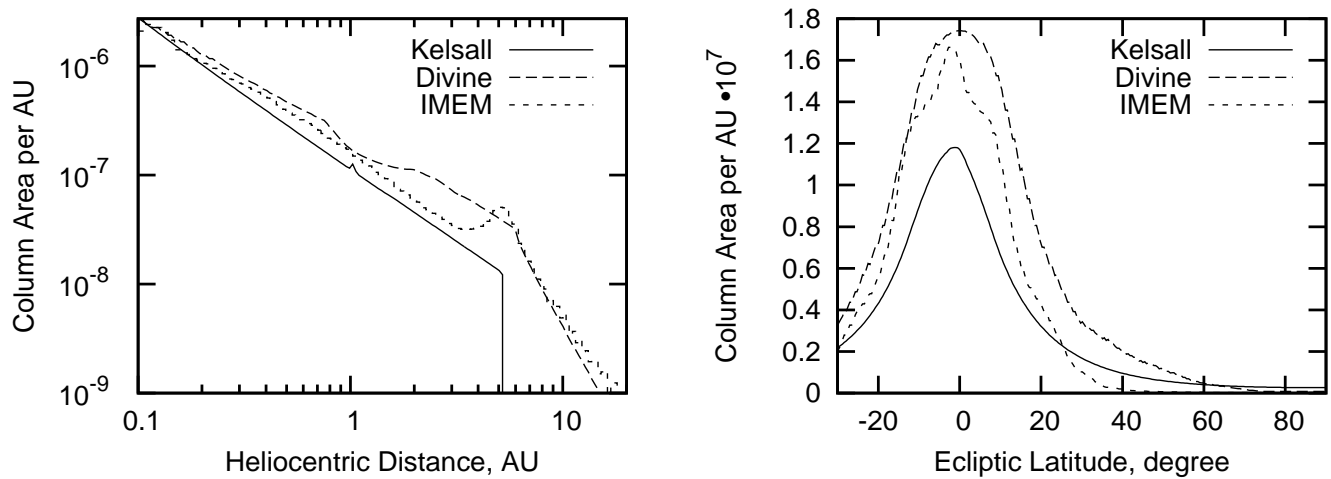}}
\caption{Profiles of particle cross-section area density predicted with
selected meteoroid models and compared with each other:
ecliptic radial (left) and latitudinal at 1~AU from the Sun (right,
the vernal equinox is at latitude~0$^\circ$).\label{Model profiles}}
\end{figure*}
In this paper, we use three recent and contemporary models of the Zodiacal
dust cloud due to \citet{Kelsall-et-al-1998}, \citet{Divine-1993} and \citet{Dikarev-et-al-2004EMP}.
The first model is constrained by the infrared observations only
from an Earth-orbiting satellite {\it COBE}. For brevity,
it is referred to as the Kelsall model hereafter.
The Kelsall model is most familiar to and most often used
by the Cosmic Microwave Background (CMB) research community
as it helps, and was designed to assess and mitigate the contamination of
the background radiation maps by the foreground radiation from interplanetary dust.
The two other models incorporate data obtained by diverse measurement techniques
and serve to predict meteoroid fluxes on spacecraft and to assess impact hazards.
Being applied in spacecraft design and analysis mostly, they are often dubbed as
meteoroid engineering models. We choose the Divine model and IMEM,
with the latter abbreviation standing for
the Interplanetary Meteoroid Engineering Model.
We also check if the most recent NASA meteoroid engineering model, MEM~\citep{McNamara-et-al-2004EMP},
can be used in our study, and explain why it cannot be.

The Kelsall model does not consider the size distribution
of particles in the Zodiacal cloud. The size distribution is
implicitly presented by the integral optical properties of the cloud.
In contrast, the meteoroid engineering models are obliged to specify
the fluxes of meteoroids for various mass, impulse,
or other level-of-damage thresholds, hence they provide
the size distribution explicitly.

Profiles of particle cross-section area density per unit volume of space
are plotted for the Kelsall and Divine models as well as IMEM in Fig.~\ref{Model profiles}.
Interestingly, the Kelsall model is sparser than both the Divine model and IMEM at most heliocentric distances.
The meteoroid engineering models predict substantially more dust than Kelsall does, especially
beyond 1~AU from the Sun: Divine's density is almost flat in the ecliptic plane between 1~and 2~AU,
exceeding Kelsall's density by a factor of~3 in the asteroid belt.
The IMEM density remains similar to that of the Kelsall model up to $\sim3$~AU,
showing a local maximum beyond the asteroid belt near 5~AU from the Sun,
in the vicinity of Jupiter's orbit.
IMEM possesses the latitudinal distribution somewhat narrower than that of the Kelsall
and Divine models.

All these distinctions stem from different observations and physics incorporated
in different models.
When describing them in subsequent subsections, we do not aim at selecting the best model.
We take all three of them instead, with the intention to ``bracket'' the more complex reality
by three different perspectives from different ``points of view''.

\subsection{The Kelsall model}\label{sec:The Kelsall model}

\begin{figure*}
\centerline{\includegraphics[width=\textwidth]{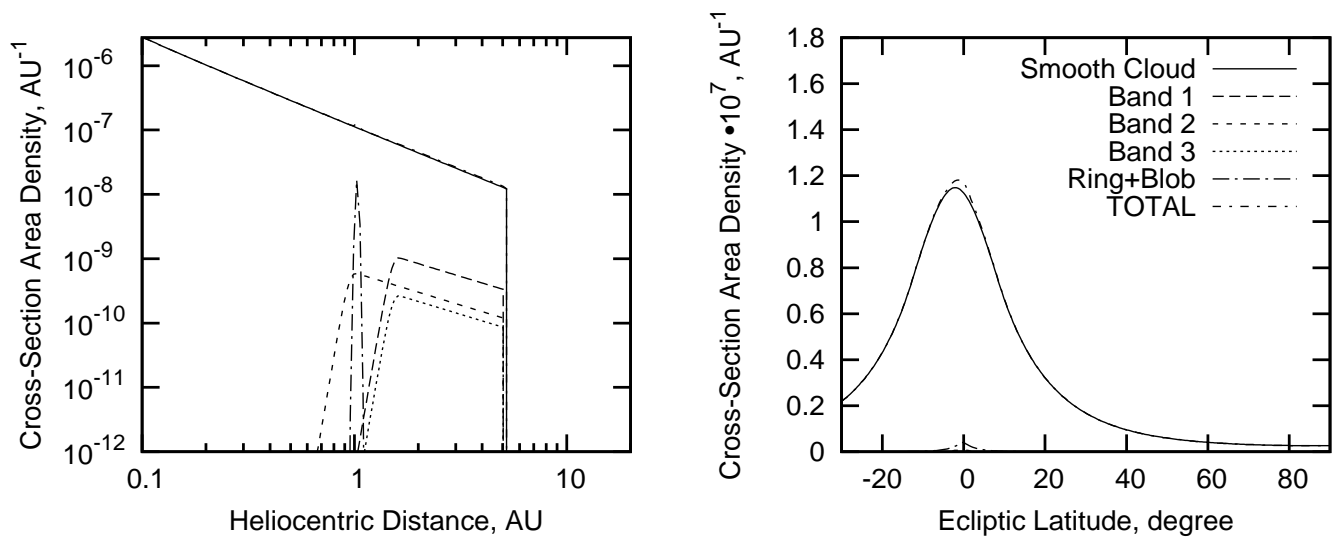}}
\caption{Radial and latitudinal profiles of particle cross-section area density
in the Kelsall model, componentwise.\label{Kelsall model componentwise}}
\end{figure*}

A concise analytical description of the infrared emission from the interplanetary dust
captured by the  {\it COBE} Diffuse InfraRed Background Experiment (DIRBE)
was proposed by \citet{Kelsall-et-al-1998}. It recognizes five emission components,
a broad and bright smooth cloud, three finer and dimmer dust bands,
and circumsolar dust ring along the Earth orbit with an embedded Earth-trailing blob.
Each component is described by a parameterised empirical three-dimensional density function
specifying the total cross-section area of the component's dust particles per unit volume of space
(Fig.~\ref{Kelsall model componentwise}),
and by their collective light scattering and emission properties
such as albedo, absorption efficiencies, etc.
The particle size distributions of the components are not considered.
Most of the light scattering and emission properties are neither adopted from
laboratory studies of natural materials nor predicted by theories of light scattering,
they are free parameters of the model fit to the DIRBE observations instead.

The smooth cloud is the primary component of the Kelsall model.
Its density function is traditionally separated into radial and vertical terms.
The radial term is a power law $1/R_c^\alpha$ with $\alpha=1.34\pm0.022$ being
the slope of the density decay with distance from the cloud's centre~$R_c$.
The slope is known to be equal to one for the dust particles in circular orbits
migrating toward the Sun under the Poynting-Robertson drag (orbiting the Sun,
the particles absorb its light coming from a slightly forward direction due to
aberration, and the radiation pressure force has a non-zero projection
against the direction of their motion; that projection causes gradual loss
of the orbital energy and decrease of the semimajor axis of particle's orbit).
The slope is greater than one if the particle orbits are initially eccentric,
or if the zodiacal cloud is fed from sources broadly distributed over 
radial distances, so that the inner circles of the cloud are supplied
from more dust sources than the outer circles~\citep{Leinert-et-al-1983,Gorkavyi-et-al-1997}.

The vertical term of smooth cloud's density is represented by a widened, modified fan model.
The smooth cloud's symmetry plane is inclined off the ecliptic plane.
This happens because the Earth has no influence on the orbital dynamics
of the vast majority of cloud's particles. The giant planets, Jupiter primarily,
perturb the orbits of dust particles as well as their sources, or parent bodies,
such as comets and asteroids, and control the inclinations of the orbits of the particles.
An offset of the center of the cloud from the Sun is also allowed in the Kelsall model
(of the order of 0.01~AU). The vertical motion of the Earth with respect to the cloud's symmetry plane leads
to the modulations of the infrared emission from the zodiacal cloud reaching 30\%
in the ecliptic polar regions.

The dust bands are the remnants of collisional disruptions of asteroids
that resulted in formation of the families of asteroids in similar orbits and remarkable dust belts.
The dust bands
were discovered first on the high-resolution maps made by the {\it IRAS} satellite
\citep{Low-et-al-1984ApJ} and recognized later in the {\it COBE} DIRBE data as well.
\citet{Kelsall-et-al-1998} introduced three dust bands in their model,
attributed to the Themis and Koronis asteroid families ($\pm1\fdg4$ ecliptic latitude),
the Eos family ($\pm10^\circ$), and the Maria/Io family ($\pm15^\circ$).
The bands have symmetry planes different from that of the smooth background cloud.
The bands are all double, with the density peaking above and below the respective symmetry plane,
since their constituent particles retain the inclination of the ancestor asteroid orbit
but the longitudes of nodes get randomized by the planetary perturbations.
The vertical motion of such particles is similar to that of an oscillator,
which moves slower and spends more time near its extremities,
thus the density at the latitudes of each family's orbital inclination is high.
One would expect a similarly shaped ``edge-brightened'' radial density with the peaks at
the perihelion and aphelion distances of the parent body orbit~\citep[e.g.\ Sect.~3 in][]{Gorkavyi-et-al-1997}.
However, the dust in asteroid bands is proven to be migrating toward the Sun
under the Poynting-Robertson drag: \citet{Reach-1992} has demonstrated that the temperatures
of particles and parallaxes of the bands measured from the moving Earth are both higher
than those supposed to be in the respective asteroid families of their origin,
implying that the bands extend farther towards the Sun than their ancestor bodies.
Unlike \citet{Reach-1992}, \citet{Kelsall-et-al-1998} allowed for a partial migration only of dust,
by introducing a cut-off at a minimal heliocentric
distance either defined or inferred individually for each band. They also fixed rather than inferred
most of the dust band shape parameters. It is noteworthy that \citet{Kelsall-et-al-1998} did not
introduce an explicit outer cut-off for the densities of the smooth cloud and dust bands.
Instead, they integrated the densities along the lines of sight from the Earth to the maximal
heliocentric distance of 5.2~AU (close to the Jovian orbital radius).

The dust particles on nearly circular orbits migrating under the Poyinting-Robertson drag
toward the Sun are temporarily trapped in a mean-motion resonance with the Earth near 1~AU \citep{Dermott-et-al-1984}.
The resulting enhancement of the zodiacal cloud density is described in the Kelsall model
by the solar ring and Earth-trailing blob. The trailing blob is the only component of the model
that is not static: it is orbiting the Sun along a circular orbit with a period of one year,
and as the name suggests, its density peak is located behind the moving Earth.
The solar ring and trailing blob have density peak distances slightly outside the Earth orbit,
their symmetry planes are inclined off the ecliptic plane, and the trailing blob orbits the Sun
at a constant velocity whereas the Earth velocity is variable due to the eccentricity of our planet's orbit.
Consequently, the Earth moves with respect to the solar ring and trailing blob over the course of the year
much like the other components of the Kelsall model.

The DIRBE instrument performed a simultaneous survey of the sky in 10 wavelength bands
at 1.25, 2.2, 3.5, 4.9, 12, 25, 60, 100, 140, and 240~$\mu$m.
It was permitted to take measurements anywhere between the solar elongations
of $64^\circ$ and $124^\circ$. As the Earth-bound {\it COBE} observatory
progressed around the Sun, DIRBE took samples of the infrared background radiation
from all over the sky. The foreground radiation due to interplanetary dust
was a variable ingredient of the sky brightness 
because it depends on the changing position of observatory with respect to the cloud.
Thus the fitting technique was based on minimizing the difference between
the brightness variations in time observed along independent lines of sight
and those predicted by the Kelsall model for the same lines of sight,
ignoring the underlying photometric baselines
contaminated or even dominated by the background sources.


The accuracy of the Kelsall model in describing the infrared emission
from interplanetary meteoroids in DIRBE's wavelength bands and
within the range of permitted solar elongations is reported to be better than 2\%.
Interpolations of the model brightnesses between the instrument wavelengths
and extrapolations beyond the wavelength and solar elongation ranges
may be prone to higher uncertainties. Indeed, the concise description of the
zodiacal emission contribution to the DIRBE data provides no clue
as to how the light scattering and emission properties of its components
depend on wavelength between and beyond the ten DIRBE wavelength bands.

When applying the Kelsall model in the far infrared wavelengths and microwaves,
one should bear in mind that already \citet{Kelsall-et-al-1998} found
the 140 and 240~$\mu$m bands nearly useless in constraining
the density distribution parameters due to relatively high detector noise
and small contribution of the radiation from the interplanetary dust at these wavelengths.
Consequently, the inferred parameters of the density distributions in the Kelsall model
are based on the observations at the wavelengths up to 100~$\mu$m.
The dust particles are efficient in interacting with
the electromagnetic radiation if their sizes are not too small
with respect to the wavelength ($2\pi s>\lambda$, where $s$ is the particle radius
and $\lambda$ is the wavelength). Dust grains with radii from $\sim16\;\mu$m are
therefore visible at the wavelength of 100~$\mu$m.
\citet{Dikarev-et-al-2009} argue that a considerable amount of meteoroids
are present in the Solar system with the sizes bigger than $\sim100\;\mu$m
which are outshined in the infrared light by more abundant and ubiquitous smaller dust grains.
The longest observation wavelength of DIRBE at which \citet{Kelsall-et-al-1998}
could still constrain the density distributions of dust is short of
being capable to resolve these particles unless they compose distinctive
features like dust bands.

We use the formulae for the density functions provided by \citet{KenGanga-2013}
since the original paper by \citet{Kelsall-et-al-1998} contains typographical misprints
in the definitions of the asteroid dust band and circumsolar ring densities.

\subsection{The Divine model}

\begin{figure*}
\centerline{\includegraphics[width=\textwidth]{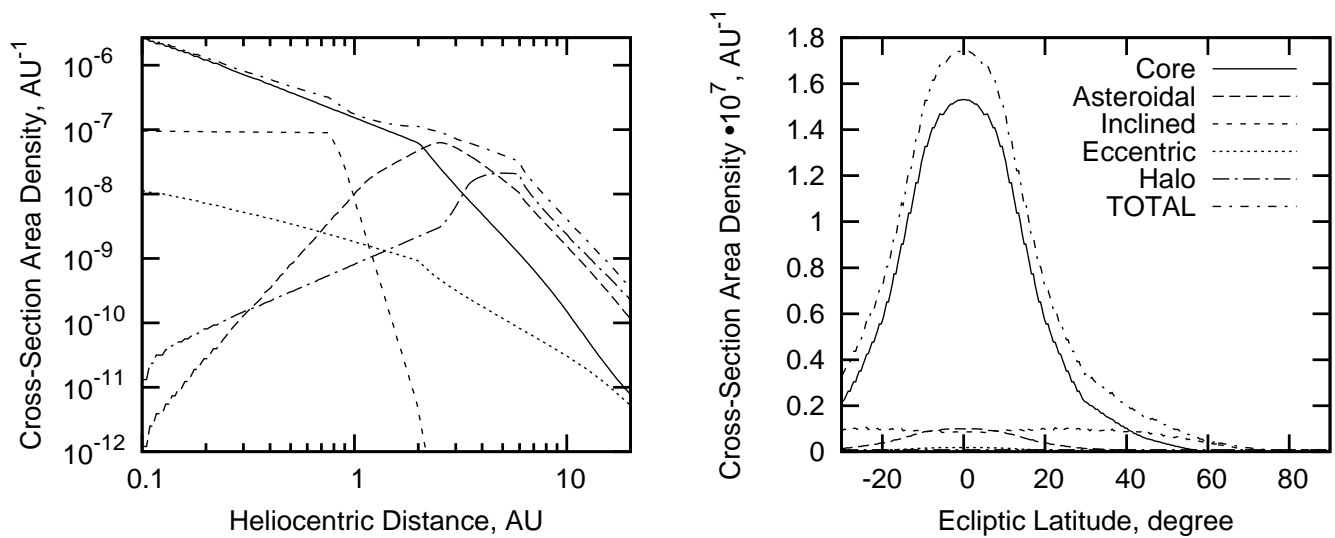}}
\caption{Radial and latitudinal profiles of particle cross-section area density
in the Divine model, componentwise.\label{Divine model componentwise}}
\end{figure*}

A model of the interplanetary meteoroid environment
\citep{Divine-1993} constructed to predict fluxes on spacecraft
anywhere in the Solar system from 0.05 to 40~AU from the Sun,
is composed of five distinct populations, each of which
has mathematically separable distributions in particle mass
and in orbital inclination, perihelion distance, and eccentricity
(Fig.~\ref{Divine model componentwise}).
Using the distributions in orbital elements,
Divine explicitly incorporated Keplerian dynamics of meteoroids
in heliocentric orbits.

The Divine model is constrained by a large number of diverse meteoroid data sets.
The mass distribution of meteoroids from $10^{-18}$ to $10^2$~g is fitted to
the interplanetary meteoroid flux model \citep{Gruen-et-al-1985}, which in turn is based on
the micro-crater counts on lunar rock samples retrieved by the {\it Apollo} missions,
i.e.\ the natural surfaces exposed to the meteoroid flux near Earth, and data from several spacecraft.
The orbital distributions are determined using meteor radar data, observations of zodiacal light
from the Earth as well as from the {\it Helios} and {\it Pioneer}~10 interplanetary probes,
and meteoroid fluxes measured in-situ by impact detectors on board {\it Pioneer}~10 and~11, {\it Helios}~1,
{\it Galileo} and {\it Ulysses} spacecraft at the heliocentric distances ranging from 0.3 to 18~AU.
The logarithms of the model-to-data ratios were minimized in a root-mean-square sense.
When modeling the zodiacal light, Divine assumed that the scattering function is independent
of meteoroid mass. The albedos of dust particles were defined somewhat
arbitrarily in order to hide the populations necessary to fit the in-situ measurements
from the zodiacal light observations. No infrared observations were used to constrain the model.

The core population is the backbone of the Divine model, it fits as much data as possible
with a single set of distributions. The distribution in pe\-ri\-he\-lion distance~$r_\pi$ of the core-population
particles can be used as strictly proportional to $r_\pi^{-1.3}$ up to 2~AU from the Sun. This function
resembles a spatial concentration, and its slope is close to Kelsall's $\alpha=1.34$ for the radial
density term: much like the infrared data from {\it COBE}, the zodiacal light observations from the Earth
and interplanetary probes demand that the slope be steeper than a unit exponent.
The eccentricities are moderate (peaking at 0.1 and mostly below 0.4),
inclinations are small (mostly below $20^\circ$).

The other populations fill in the gaps where one population with separable distributions
is not sufficient to reproduce the observations. Their names hint at distinctive features
of their orbital distributions. The inclined population is characterized by inclinations largely in the range
$20^\circ$--$60^\circ$, its eccentricities are all below 0.15. In contrast, the eccentric population
is composed of particles in highly eccentric orbits (the eccentricities are largely between 0.8 and 0.9),
the inclinations are same as in the core population. These two populations are
added in order to compensate for a deficit of meteoroids from the core population with respect
to in-situ flux measurements on board the {\it Helios} spacecraft. (The eccentric population is also used
to match the interplanetary flux model between $10^{-18}$ and $10^{-15}$~g.)
The halo population has a uniform distribution of orbital plane orientations
and ``surrounds'' the Sun as a halo between roughly 2 and 20~AU.
It patches the meteoroid model where the above-listed populations
are not sufficient to reproduce the {\it Ulysses} and {\it Pioneer} in-situ flux measurements.
The eccentric, inclined and halo populations are composed of meteoroids much smaller
than those significant enough for predictions of the infrared and microwave observations.

The asteroidal population is described last but it makes a huge difference between Kelsall's and Divine's models.
Already when fitting the first population to the meteor radar data, Divine observed that the results for
the distribution in perihelion distance show a minimum near $r_\pi=0.6$~AU,
suggesting that two components are involved (i.e., inner and outer). The radial slope
of the inner component was consistent with that demanded by the zodiacal light observations, and
the inner and outer components contributed 45\% and 55\%, respectively, to the flux near Earth.
The outer component's concentration peaked outside the Earth orbit, in the asteroid belt.
Since the main contribution to the zodiacal light comes from the particles with masses smaller than
$10^{-5}$~g and the median meteoroid mass of the meteor radar was $10^{-4}$~g, it was reasonable to ascribe the
inner fraction to a population dominated by smaller particles responsible for the zodiacal light,
and the outer fraction to another one dominated by larger particles detected as meteors only with the radar.
The two fractions were divided into the core and asteroidal populations, accordingly. Their
inclination and eccentricity distributions are identical, but the distributions in perihelion distance
and mass are different. The asteroidal population is used to fit the interplanetary flux model
at large masses $>10^{-4}$~g.

Derived from data analysis rather than postulated theoretically, the distinction between the core
and asteroidal populations in the Divine model separates -- both on the mass and
perihelion distance scales -- the small dust particles migrating toward the Sun under the Poynting-Robertson drag,
the concentration of which increases with the decrease of the heliocentric distance,
and their immediate parent bodies, the larger particles swarming further away from the Sun,
i.e.\ in the asteroid belt according to \citet{Divine-1993}.



\subsection{IMEM}

\begin{figure*}
\centerline{\includegraphics[width=\textwidth]{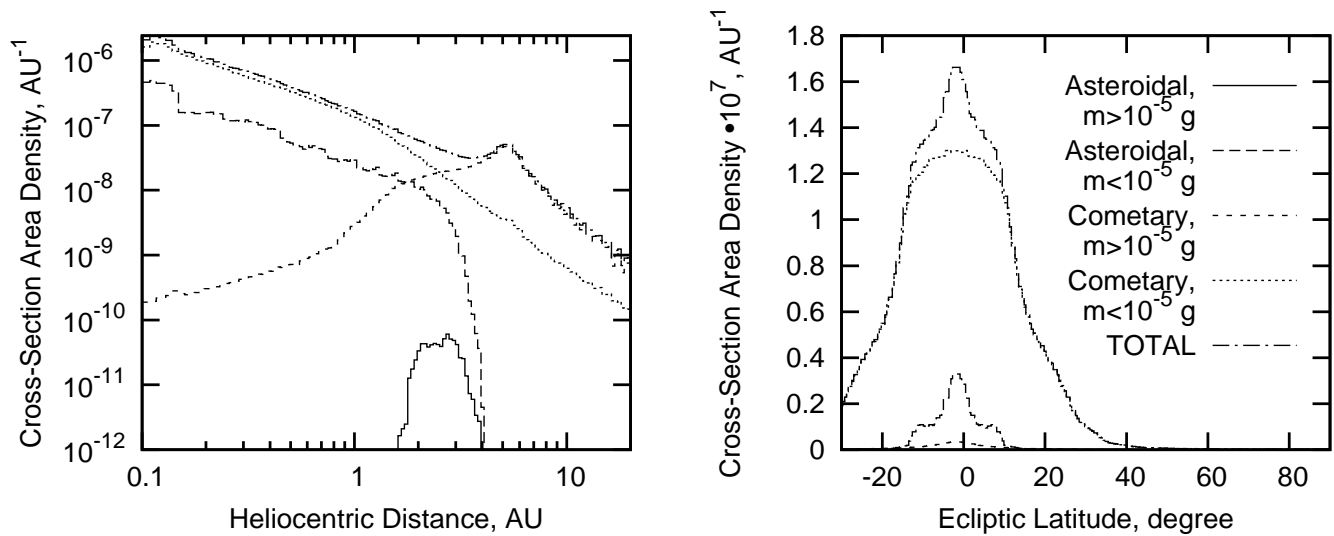}}
\caption{Radial and latitudinal profiles of particle cross-section area density
in IMEM, componentwise. The density plots are not smooth since the orbital
distributions in IMEM are approximated by step functions.\label{IMEM componentwise}}
\end{figure*}
The In\-ter\-pla\-ne\-ta\-ry Me\-teo\-roid En\-gi\-nee\-ring Mo\-del (IMEM) is developed
for ESA by \citet{Dikarev-et-al-2004EMP}. Like the Divine model, it uses the distributions
in orbital elements and mass rather than the spatial density functions of the Kelsall model,
ensuring that the dust densities and fluxes are predicted in accord with Keplerian dynamics
of the constituent particles in heliocentric orbits.
Expansion of computer memory has also enabled the authors of IMEM
to work with large arrays representing multi-dimensional distributions in orbital elements
to replace Divine's separable distributions. The distribution in mass remains separable
from the 3-D distribution in inclination, perihelion distance and eccentricity.

IMEM is constrained by the micro-crater size statistics collected from the lunar rocks
retrieved by the {\it Apollo} crews,
{\it COBE} DIRBE observations of the infrared emission from the interplanetary dust at
4.9, 12, 25, 60, and 100~$\mu$m wavelengths, and {\it Galileo} and {\it Ulysses} in-situ
flux measurements. An attempt to incorporate new meteor radar data from the Advanced
Meteor Orbit Radar \citep[{\it AMOR}, see][]{Galligan-Baggaley-2004} was not successful \citep{Dikarev-et-al-2005},
as the latitudinal number density profile of meteoroids derived from the
meteor radar data stood in stark contrast to that admissible by the {\it COBE} DIRBE data,
due to incomplete understanding of the observation biases.

The meteoroids in IMEM are distributed in orbital elements in accord with an approximate
physical model of the origin and orbital evolution of particulate matter under the planetary gravity,
Poynting-Robertson effect and mutual collisions. \citet{Gruen-Dikarev-2009LanB} visualise
the IMEM distributions of meteoroids in mass and in orbital elements in graph and color plots.
Cross-section area density profiles are shown here in Fig.~\ref{IMEM componentwise}.

\begin{figure}
\begin{center}
\includegraphics[width=0.775\textwidth,angle=270]{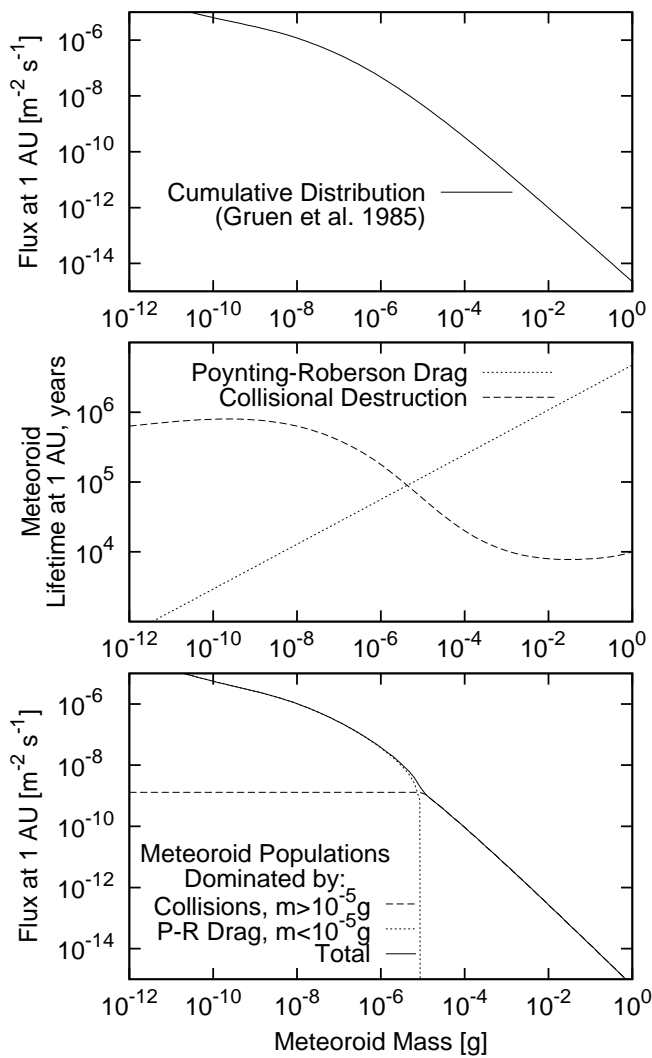}
\end{center}
\vspace{-11pt}
\caption{Cumulative mass distribution of interplanetary meteoroids \citep[top]{Gruen-et-al-1985},
meteoroid lifetimes against mutual collisions and Poynting-Robertson drag (middle), and
cumulative mass distributions of meteoroid populations in IMEM~\citep[bottom]{Dikarev-et-al-2005}.
The flux of meteoroids with $m>10^{-5}$~g is lower in IMEM than in the Gr\"un model
since the impact velocities of big meteoroids are higher at 1~AU from the Sun in IMEM
and they produce larger craters on the lunar rocks than in the Gr\"un model,
which assumed a single impact velocity for all meteoroid sizes.\label{flux-life-pops at 1 AU}}
\end{figure}

All interplanetary meteoroids are divided into populations by origin/source and mass/dynamical regime.
\citet{Gruen-et-al-1985} constructed a model of the mass distribution of interplanetary meteoroids.
Based on this model, they estimated particle lifetimes against two destructive processes,
mutual collisions and migration downward toward the Sun due to the Poynting-Robertson drag (terminated
by particle evaporation or thermal break-up). Figure~\ref{flux-life-pops at 1 AU} shows that
the lifetime against mutual collisions is shorter than the Poynting-Robertson time
for the meteoroids bigger than $\sim10^{-5}$~g in mass, located at 1~AU from the Sun.
The destruction of meteoroids is dominated by a factor of ten in rate
by the respective process already one order of magnitude away from this transition mass in either direction.
The rates vary with the distance from the Sun, however, the transition mass remains nearly unchanged
\citep[Fig. 6 in][]{Gruen-et-al-1985}. That is why \citet{Dikarev-et-al-2004EMP} could simplify
the problem by dividing the mass range into two subranges with distinct mass distribution slopes and
dynamics. Meteoroids below the transition mass of $10^{-5}$~g are treated as migrating from the
sources toward the Sun under the Poynting-Robertson drag, meteoroids above the transition mass
are assumed to retain the orbits of their ultimate ancestor bodies over their short lifetimes until
collisional destruction.

The ultimate ancestor bodies of interplanetary dust in IMEM are comets and asteroids. A catalogue
of asteroids is used to account for the latter source of meteoroids. Three distinct populations
of asteroids are recognized and allowed to have independent total production rates:
the Themis and Koronis families (semi-major axes $2.8<a<3.25$~AU, eccentricities $0<e<0.2$, and inclinations
$0<i<3\fdg5$), Eos and Veritas families ($2.95<a<3.05$~AU, $0.05<e<0.15$, $8\fdg5<i<11\fdg5$), and
the main belt ($a<2.8$~AU). The orbital distributions of meteoroids more than $10^{-5}$~g in mass are
constructed by counting the numbers of catalogued asteroids per orbital space bins. The orbital
distributions of meteoroids less than $10^{-5}$~g in mass are described as the flow of particles along
the Poynting-Robertson evolutionary paths~\citep{Gorkavyi-et-al-1997} starting from the source distributions
defined earlier. The mass distributions are adopted piecewise from the interplanetary meteoroid flux model
by \citet{Gruen-et-al-1985} as shown in Fig.~\ref{flux-life-pops at 1 AU}.

The orbital distributions of meteoroids from comets cannot be defined as easy as those of meteoroids
from asteroids. Because of a number of loss mechanisms, such as comet nuclei decay, accretion,
tidal disruption or ejection from the Solar system by planets,
very few comets are active at a given epoch and listed in the catalogues. Moreover,
the catalogues are prone to observation biases since the comet nuclei are revealed by gas and dust
shed at higher rates at lower perihelion distances.

In order to describe the orbital distributions of meteoroids of cometary origin, \citet{Dikarev-Gruen-2004ASP}
proposed an approximate analytical solution of the problem of a steady-state distribution of particles
in orbits with frequent close encounters with a planet.
This solution is applied to represent the orbital distributions of meteoroids
from comets in Jupiter-crossing orbits, i.e.\ the vast majority of catalogued comets.
Big meteoroids with masses greater than $10^{-5}$~g are confined to the region of close
encounters with Jupiter.
There is only one quantity that is approximately conserved in the region of close encounters with Jupiter.
It is the Tisserand parameter~$T$
$$
   T={\frac{a_\mathrm{J}}{a}}+2\sqrt{{\frac{a}{a_\mathrm{J}}}(1-e^2)}\cos i,
$$
with $a_\mathrm{J}$ being the semimajor axis of Jovian orbit.
The numbers of meteoroids in $T$-layers are therefore stable over a long period of time.
They are free parameters to be determined from the fit of IMEM to observations.
The relative distributions in orbital elements within each $T$-layer are found
theoretically~\citep{Dikarev-Gruen-2004ASP}.

Finer dust grains with masses less than $10^{-5}$~g leak from
this region into the inner solar system and then migrate toward the Sun under the Poynting-Robertson drag.
The Poynting-Robertson migration time is much longer than the close approach time in the region of close encounters with Jupiter,
consequently, the orbital distributions of all meteoroids are shaped there by the gravitational scattering on Jupiter.
It is only the dust leaking through the inner boundary of the region that
is distributed as the flow of particles along the Poynting-Robertson evolutionary paths~\citep{Gorkavyi-et-al-1997}.
Their distributions are also established theoretically in IMEM, with the normalisation factors
being inferred from the fit to observations.

An empirical finding by Divine of a helpful separation of the bulk of meteoroid cloud
into the core and asteroidal populations -- segregation of big and small dust particles -- has
become one of the physical assumptions in IMEM.

The weights of populations of dust from comets and asteroids in IMEM were fitted to in-situ flux measurements
using the Poisson maximum-likelihood estimator and to infrared observations using the Gaussian
maxiumum-likelihood estimator.

IMEM was tested against those data not incorporated in the model by \citet{Dikarev-et-al-2005},
confirming that the orbital evolution approach allows for more reliable extrapolations
of the observations and measurements incorporated in the model,
it is compared with several other meteoroid models by \citet{Drolshagen-et-al-2008EMP}.

\subsection{MEM}

The Meteoroid Engineering Model (MEM)
described by \citet{McNamara-et-al-2004EMP} is another recent development
of the orbital distributions and software to predict fluxes on spacecraft
in the Solar system and near Earth.
It is constrained by the Earth-based meteor radar data {\it CMOR}
(Canadian Meteor Orbit Radar)
and zodiacal light observations from two interplanetary probes, {\it Helios}~1 and~2.
The nominal heliocentric distance range at which the model is applicable,
from 0.2 to 2~AU, is rather limited, however.
Even though the meteoroid mass range of 10$^{-6}$~to 10~g
covered by the model is extremely interesting for the purposes
of our study, some assumptions made by the authors are rather
arguable. In particular, the mutual collisions between
meteoroids are considered in the model ignoring
dependence of the meteoroid collision probability
on its size, whereas \citet{Gruen-et-al-1985} calculated
that the lifetime against collisional disruption of meteoroids
varies by a factor of 100 between the masses of $10^{-6}$ and 1~g (their Fig.~6 and
our Fig.~\ref{flux-life-pops at 1 AU} in the text above)! It is not only the mass distribution of particles that is
determined by the collision probability, but also the orbital
distributions: as the lifetime against collisions becomes
shorter than the Poynting-Robertson time by orders of magnitude,
the meteoroids can no longer survive a travel
from their sources toward the Sun and the orbital
distributions of bigger particles are more similar with the distributions
of their sources than the distributions of smaller particles.
MEM is unfortunately unable to reveal and describe this important feature
of the interplanetary meteoroid cloud.

\subsection{The Cross-Section Area Distributions}

\begin{figure*}
\centerline{\includegraphics[width=0.9\textwidth]{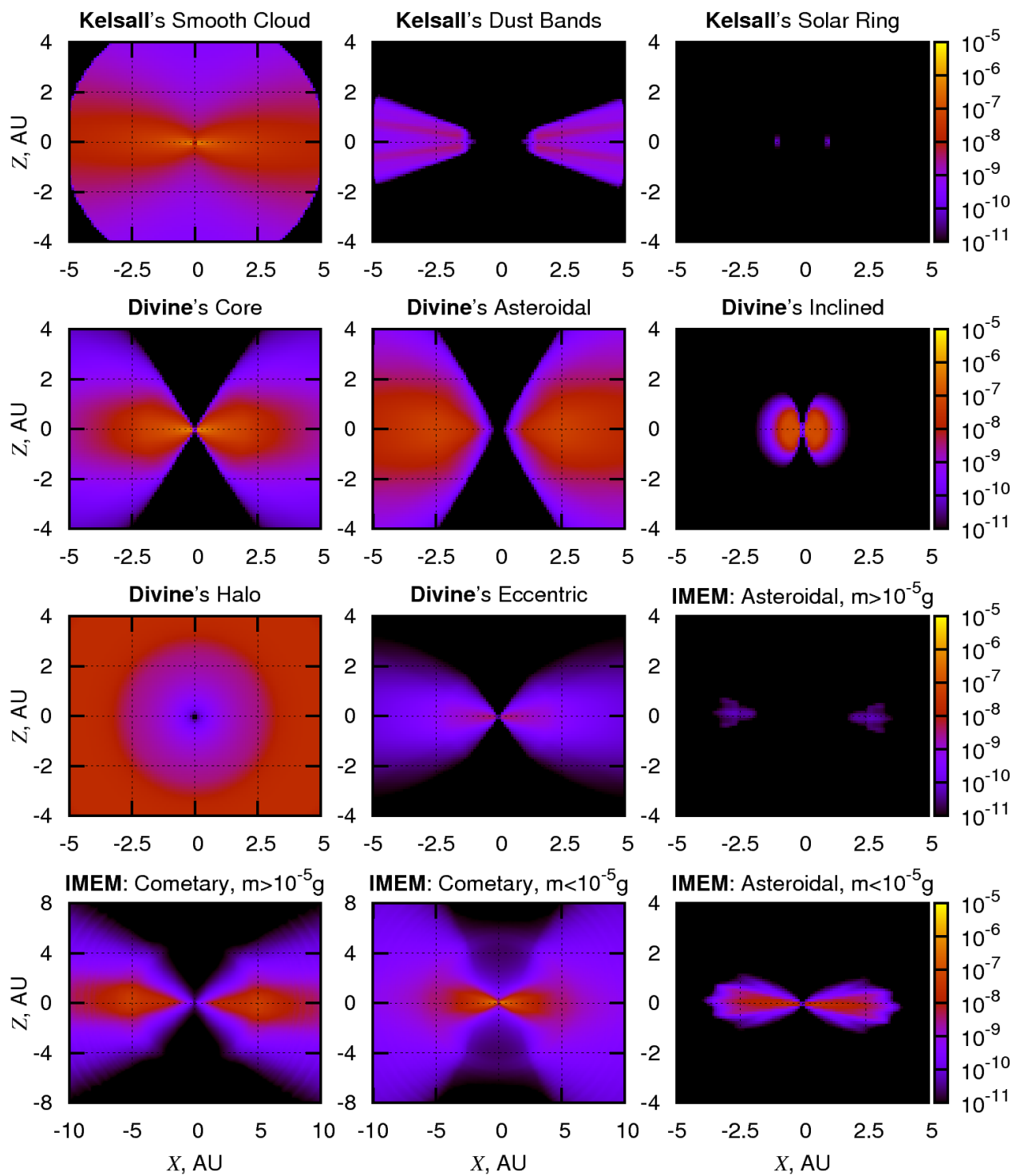}}
\caption{Particle cross-section area densities (per unit volume of space, AU$^2$/AU$^3$)
for each component or population of the meteoroid models by \citet{Kelsall-et-al-1998},
\citet{Divine-1993} and IMEM~\citep{Dikarev-et-al-2004EMP}.
Note that two maps for the cometary dust in IMEM are plotted in different scale.\label{Model sections}}
\end{figure*}
Figure~\ref{Model sections} maps the total cross-section area of dust particles
per unit volume of space, as a function of the position in the Solar system, for each component
or population of every meteoroid model introduced in the previous Section.
The~$X$ and~$Z$ coordinates are measured from the Sun,
with the $X$~axis pointing to the vernal equinox
and $Z$~axis being perpendicular to the ecliptic plane.
The integral of the cross-section area density along a line of sight gives
the optical depth of the cloud along that line of sight,
assuming the geometric-optics approximation, spherical particles, and ignoring that
the particle efficiencies in absorbing and scattering light can in fact
be higher or lower than unity.

The Kelsall model is reproduced in the upper row of maps. Three dust bands are
shown together in the middle plot. The smooth cloud is to the left,
and the solar ring is to the right of it. The dominant component of the Kelsall model,
the smooth cloud is getting higher in density only toward the Sun and the symmetry plane slightly
inclined off the ecliptic plane. The dust bands are the only component of the model
bulking beyond the Earth orbit, and they are by definition bound to the ecliptic latitudes
of their ancestor asteroid families' orbit inclinations. Both the smooth cloud and the dust bands
extend up to the heliocentric distance of 5.2~AU only, since \citet{Kelsall-et-al-1998} stopped
integrating the model densities there. A more rigorous model of the bands composed
of dust migrating toward the Sun due to the Poynting-Robertson effect would put
their densities' cut-offs at the outer boundaries of the corresponding asteroid
families. Note, however, that most of the thermal infrared emission from the
interplanetary dust observed by {\em COBE} was due to the dust within 0.5--1~AU from the Earth orbit,
since {\em COBE} could not be pointed too close to the Sun and dust is
too cold and inefficient at thermal emission far from the Sun~\citep[their Sect.~2.1 and Fig.~2]{Dikarev-et-al-2009}.
Thus for the purpose of modelling the infrared thermal emission
observed from the Earth, the behaviour of the density at multiples
of an astronomical unit was not important. This may not be the case
for the microwave emission though.

The next two rows of maps depict the populations of the Divine model. The core and
asteroidal populations are important in the infrared and microwave emission ranges.
They are distributed remarkably different in space, with the core population density
growing higher toward the Sun and the asteroidal population density peaking
beyond the orbit of the Earth. The two populations are also composed of particles
of different sizes: the core population mostly smaller and the asteroidal population
mosly bigger than $\sim50\;\mu$m. The halo, inclined and eccentric populations
are provided for the sake of completeness. Their tiny, up to $\sim10\;\mu$m-sized
particles are ignorable in the wavelength ranges of interest.

The bottom row of maps and the right map in the third row exhibit the populations from IMEM.
The asteroidal dust particles bigger than $10^{-5}$~g in mass are confined
to the asteroid belt, naturally. They are not migrating from their ancestor
body orbits toward the Sun nor extend beyond the outer edge of the asteroid belt.
The smaller dust particles move their ways toward the Sun due to the Poyting-Robertson
effect. The inclinations of their orbits are intact, since IMEM
does not take into account any planetary perturbations other than gravitational
scattering by Jupiter that require a close approach to the giant planet.
Consequently, the latitudinal distribution of their density is independent
on heliocentric distance away from the source region, i.e.\ asteroid belt.

\begin{figure*}
\begin{center}
\includegraphics[width=\textwidth]{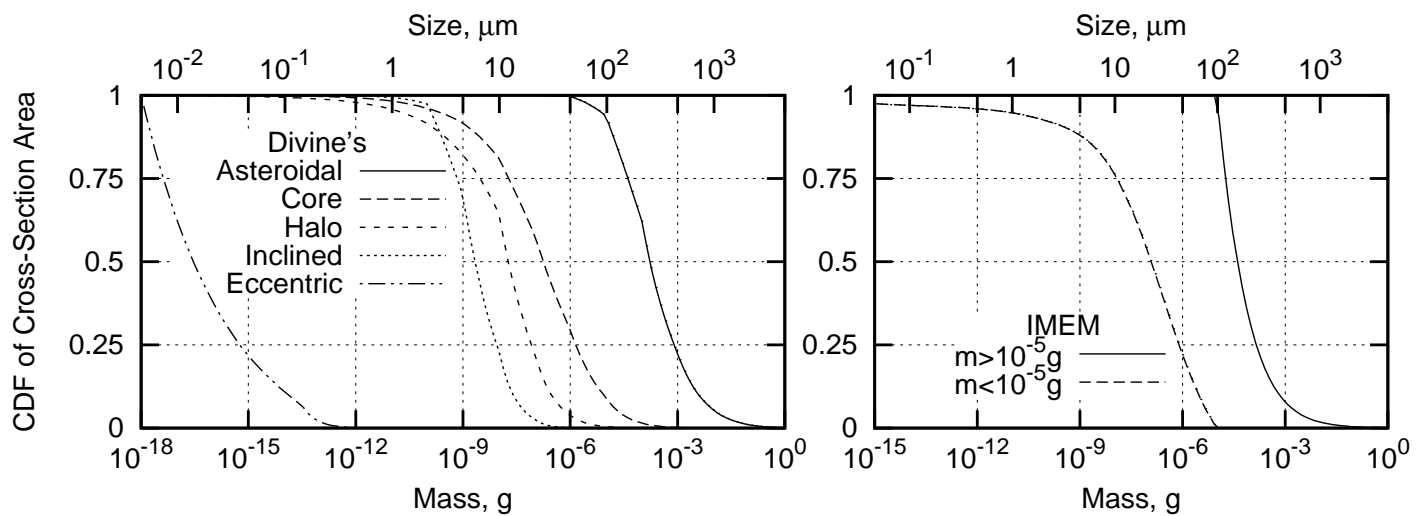}
\end{center}
\caption{Cumulative distribution functions of particle cross-section area in mass (size)
for each population of the interplanetary meteoroid model by \citet{Divine-1993}
and IMEM~\citep{Dikarev-et-al-2004EMP}. The material density of particles
is equal to 2.5~g~cm$^{-3}$ in both models except Divine's eccentric population,
in which it is ten times lower. The particle size scale should not be used
in combination with the plot for the latter population.\label{Cross-section}}
\end{figure*}
Big cometary particles have a density peak along the orbit of Jupiter. This occurs
because all their orbits are required to approach Jovian orbit within 0.5~AU or less.
Small cometary particles leak through the inner boundary of the region of close
encounters with Jupiter and migrate toward the Sun under the action of the Poynting-Robertson
effect, they reach the highest density at the shortest heliocentric distances.
A cometary dust density enhancement in the form of a spherical shell with a radius of 5.2~AU
is a negligibly small defect caused by the assumption of a uniform distribution
of particle orbits in longitudes of perihelia in IMEM. Concentric spherical ``shells''
of different densities on the map of cometary meteoroids with $m>10^{-5}$~g are
due to finite discretization of the distribution in perihelion distance.

Cumulative distribution functions of meteoroid cross-section area in mass and size for each
population of the meteoroid models by \citet{Divine-1993} and \citet{Dikarev-et-al-2004EMP}
are plotted in Fig.~\ref{Cross-section} (normalized to unity). Note that one population
of the Divine model has an assumed material density of 0.25~g~cm$^{-3}$, whereas the standard value
assumed for all other populations as well as in many other meteoroid models is 2.5~g~cm$^{-3}$.

The Kelsall model bets on essentially single-population representaion of the
interplanetary meteoroid cloud, with the solar ring and dust bands being
rather minor features (Fig.~\ref{Kelsall model componentwise}).
The meteoroid engineering models, however,
state it clearly that the cloud is composed of different sorts of dust
in major populations that are also distributed differently in space.
We will see below how this affects the microwave emission predictions.

\section{Thermal Emission Models}\label{Thermal Emission Models}

The thermal emission from dust particles is expressed
with respect to the blackbody emission~$B_\lambda(T)$
at the wavelength~$\lambda$ and temperature~$T$,
using an emissivity modification factor
which matches the absorption efficiency factor~$Q_\mathrm{abs}(s,\lambda)$
for the particle size~$s$~\citep{Bohren-Huffman-1983}.
$Q_\mathrm{abs}$ is defined as the ratio of the effective absorption cross-section
area of the particle to its geometric cross-section area.

\subsection{The Kelsall model}\label{sec:Kelsall thermal}

\begin{figure*}
\begin{center}
\includegraphics[width=\textwidth]{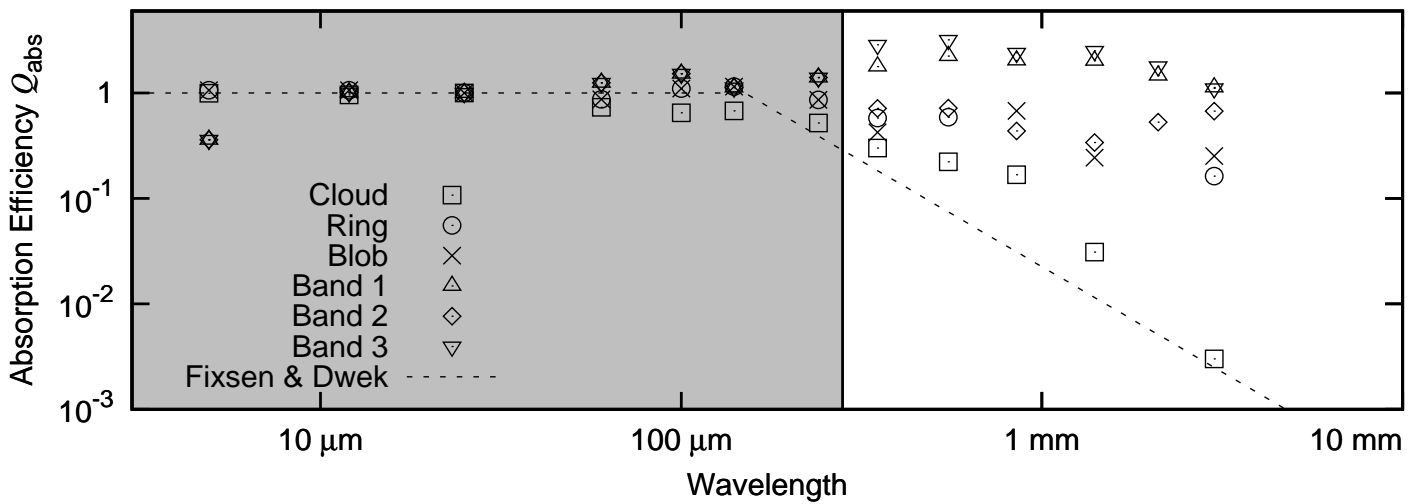}
\end{center}
\caption{Absorption efficiencies of dust in the Kelsall model determined
for the {\it COBE} DIRBE wavelengths by \citet{Kelsall-et-al-1998},
gray area on the plot, and for the {\it Planck}/HFI wavelengths
by \citet{KenGanga-2013}. \citet{Fixsen-Dwek-2002} have also used
the data from {\it COBE}/FIRAS instrument
to determine the annually averaged spectrum of the Zodiacal cloud
plotted here with a dotted line.\label{Emi-COBE-Planck}}
\end{figure*}
The light scattering and emission properties of dust are not
defined in the Kelsall model for the wavelengths between and beyond
those of the {\it COBE} DIRBE instrument. Figure~\ref{Emi-COBE-Planck} shows Kelsall's
emissivity modification factors, or absorption efficiencies to
preserve the uniformity of terms, for the DIRBE wavelength bands. \citet{KenGanga-2013}
used the cross-section area density of the Kelsall model and
found the absorption efficiencies for its components
to describe approximately the thermal emission from interplanetary dust
at the wavelengths of {\it Planck's} High Frequency Instrument (HFI).
\citet{Fixsen-Dwek-2002} used the FIRAS (Far-Infrared Absolute Spectrometer) data from {\it COBE} and
found that the annually averaged spectrum of the zodiacal cloud
can be fitted with a single blackbody at a temperature
of 240~K with an absorption efficiency being flat
at wavelengths shorter than 150~$\mu$m and $Q_\mathrm{abs}\propto\lambda^{-2}$
beyond 150~$\mu$m.

The absorption efficiencies of dust from all three bands coincide in \citet{Kelsall-et-al-1998}
for the wavelengths up to 240~$\mu$m. \citet{KenGanga-2013} have removed this constraint
and allowed for individual weights of the band contributions in the microwaves.
They found that the bands keep high~$Q_\mathrm{abs}\sim1$ up to~$\lambda\sim3$~mm,
implying that their constituent particles are macroscopic.
The absorption efficiency of the smooth cloud decays in the microwaves,
however, not exactly as sharp as a simple approximation of \citet{Fixsen-Dwek-2002} suggests.

The inverse problem solution sometimes led \citet{KenGanga-2013} to negative absorption efficiencies
of the smooth cloud or circumsolar dust ring and Earth-trailing blob.
(Those negative efficiencies are simply missing from Fig.~\ref{Emi-COBE-Planck}
at certain wavelengths, as the logarithmic scale does not permit negative ordinates.)
Obviously, some components of the Kelsall model were used by the fitting procedure to compensate for
excessive contributions from the other components in such cases. This is a strong indication of
insufficiency of the Kelsall model in the microwaves.

Our study requires the absorption factors even further in the microwaves.
We use the numbers inferred by \citet{Kelsall-et-al-1998} and \citet{KenGanga-2013} whenever
possible, i.e. when they are available and positive. A negative absorption efficiency
found for the smooth cloud by \citet{KenGanga-2013} at $\lambda=2.1$~mm is replaced
with the result of interpolation of the positive efficiencies found at $\lambda=1.4$ and 3~mm,
whereas negative efficiencies for the circumsolar ring at $\lambda=0.85$, 1.4 and 2.1~mm
are simply nullified. The absorption efficiency is extrapolated beyond the {\it Planck}/HFI
wavelengths ($\lambda>3$~mm, i.e.\ in the {\it WMAP} range) using the approximation due to
\citet{Fixsen-Dwek-2002} for the smooth cloud~$Q_\mathrm{abs}\propto\lambda^{-2}$,
similar to \citet{Maris-et-al-2006AandA} who assessed the level of contamination of the {\it Planck} data
by the Zodiacal microwave emission based on the Kelsall model,
and using flat $Q_\mathrm{abs}=1$, 0.5, 1, and 0.1 for the asteroid dust bands~1, 2, 3,
and the circumsolar ring with the trailing blob, respectively.

\subsection{Selected Materials and Absorption Efficiencies}

Predictions of the thermal emission from interplanetary dust clouds using the meteoroid
engineering models require optical properties of constituting particles.
Following \citet{Dikarev-et-al-2009}, we use the Mie light-scattering theory
and optical constants of silicate and amorphous carbonaceous spherical particles
in order to estimate the absorption of solar radiation and thermal emission
by meteoroids. \citet{Dikarev-et-al-2009} banned iron from substances for a hypothetical meteoroid
cloud that could be responsible for anomalous CMB multipoles. In this paper, however,
we take it back into account since iron is known to compose some meteorites
as well as asteroid surfaces, it is present in the interplanetary dust particles
and therefore it must be considered when discussing the real, not hypothetical
Solar-system medium. Water and other ices are still ignored as extremely inefficient
emitters of the microwaves. 

\begin{figure*}[t]
\begin{flushleft}
\includegraphics[width=0.525\textwidth,angle=270]{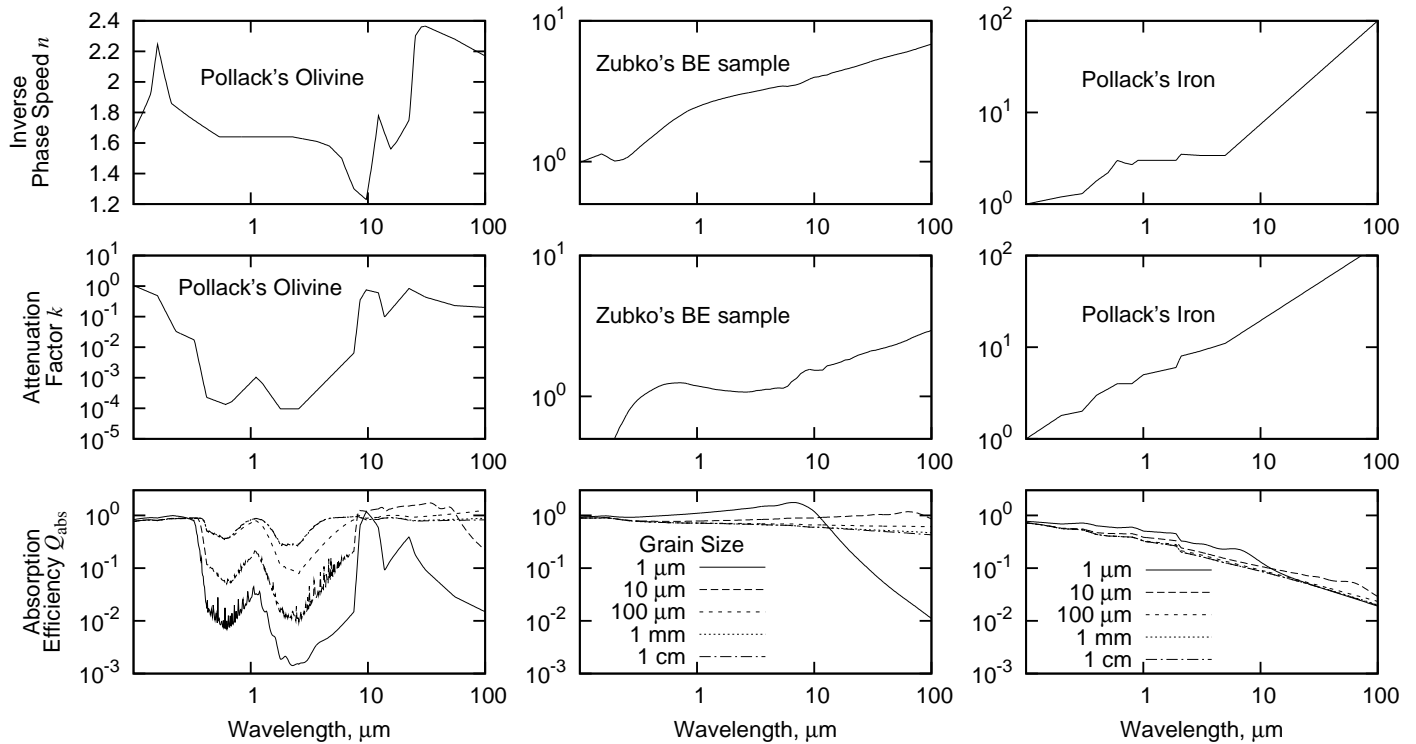}
\end{flushleft}
\caption{Optical constants and absorption efficiencies for olivine and iron of
\citet{Pollack-et-al-1994}, and for amorphous carbonaceous grains from the `BE' sample
of \citet{Zubko-et-al-1996} used in this paper, in addition to the data introduced
by \citet{Dikarev-et-al-2009}. The absorption efficiencies are calculated for the
sizes of spherical dust grains indicated in the bottom row of plots.\label{nkQabs}}
\end{figure*}

Dealing with the microwave thermal emission from dust with an assumed temperature,
\citet{Dikarev-et-al-2009} did not need optical constants for the visual and near-infrared
wavelengths. Here we calculate the temperatures of dust in thermal equilibrium and
the absorption efficiencies are required for the brightest part of the Solar spectrum. Figure~\ref{nkQabs}
plots the data for the wavelengths between~0.1 and~100$\;\mu$m.

\subsection{The Dust Temperatures}

\begin{figure}[t]
\begin{center}
\includegraphics[width=\columnwidth]{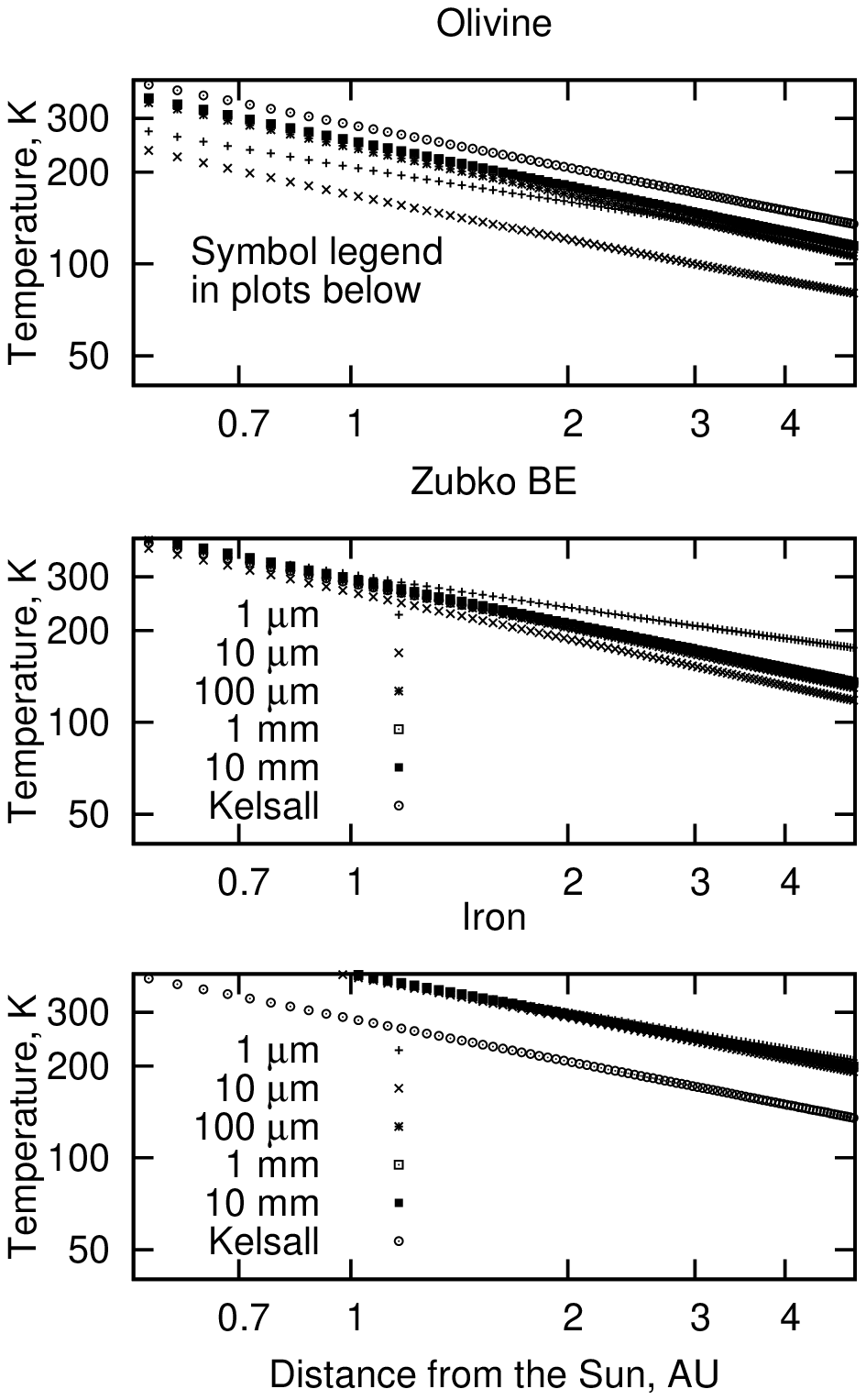}
\end{center}
\caption{Equilibrium temperatures of homogeneous spherical dust particles composed of different substances.\label{Equitemp-vs-distance}}
\end{figure}

Let us now calculate and discuss the temperatures of dust particles of different sizes
at different distances from the Sun for the substances selected in the previous section.
Figure~\ref{Equitemp-vs-distance} plots the equilibrium temperatures of spherical homogeneous
particles composed of silicate, carbonaceous and iron materials as well as
the dust temperature used in the Kelsall model.

The equilibrium temperature is found from the thermal balance equation
\begin{equation}
   \pi s^2 \int_0^\infty Q_\mathrm{abs} (s, \lambda) {\cal F}_\odot(\lambda) {\;\rm d}\lambda = 
   4\pi s^2 \int_0^\infty Q_\mathrm{abs} (s, \lambda) B_\lambda (T_\mathrm{D}){\;\rm d}\lambda,
\label{Thermal Equ}
\end{equation}
where $s$ is the radius of a spherical dust grain,
$\lambda$ is the wavelength, ${\cal F}_\odot$ is the incident Solar radiance flux,
$B_\lambda (T_\mathrm{D})$ is the blackbody radiance at the dust particle's temperature~$T_\mathrm{D}$.
As the left-hand side provides the total energy absorbed from a mono-directional
incident flux, the right-hand side gives the total energy emitted, omni-directionally.
By denoting the absorption efficiency averaged over the Solar spectrum
with~$\bar Q_\odot$, and the same quantity averaged over a blackbody spectrum
at temperature~$T_\mathrm{D}$ with~$\bar Q(T_\mathrm{D})$, then
using the Stefan-Boltzmann law and the Solar constant,
one can rewrite Eq.~(\ref{Thermal Equ}) in a more concise form~\citep[cf.][]{Reach-1988}:
\begin{equation}\label{Reach Equ}
   T_\mathrm{D} = 279\;\mathrm{K}\; \left[\bar Q_\odot / \bar Q (T_\mathrm{D})\right]^{1/4} \left({\frac{R}{1\;\mbox{AU}}}\right)^{-1/2},
\end{equation}
where $R$ is the distance from the Sun. A perfect black body with~$Q_\mathrm{abs}=1$ throughout the spectrum
has therefore a temperature of 279~K at 1~AU from the Sun, inversely proportional
to the square root of the distance. This inverse-square-root trend is often closely
followed by the real dust particles.

The Kelsall model uses the temperature given by equation
\begin{equation}
 T = 286\;\mbox{K} \left({\frac{1\;\mbox{AU}}{R}}\right)^{0.467}.
\end{equation}
\citet{Kelsall-et-al-1998} emphasize that the dust temperature at 1~AU and absorption
efficiencies could not be determined independently, so that the temperature was found
by assuming the smooth cloud to be the dominant component with its spectrum in the mid-IR
being that of a pure blackbody (i.e., unit absorption efficiencies at $\lambda=4.9$, 12, and 25~$\mu$m).

The equilibrium temperatures of dust particles are shown in Fig.~\ref{Equitemp-vs-distance}
for a broad range of heliocentric distances. Temperatures used in the Kelsall model are plotted for comparison as well. 
Temperatures of the micrometer-sized particles are in most obvious disaccord with Kelsall's model, but they
tend to be higher than temperatures of the larger particles of their composition as well.
The reason is simple: their absorption efficiencies are too low at the wavelengths of ten
and more micrometers, at which their larger counterparts emit the energy absorbed from the Solar radiation flux,
thus they warm to higher temperatures in order to shift the maximum of the Planck function
to shorter wavelengths. Iron is an exception from this rule, all particles made of this substance
are considerably warmer, being unable to emit the radiation thermally at longer wavelengths regardless of size.
(It does not contradict to the fact that communications in radiowaves are assisted by iron and other
metallic antennas, as iron is very good at scattering the electromagnetic emission at longer wavelengths simultaneously!)

\begin{figure}[t]
\begin{center}
\includegraphics[width=\columnwidth]{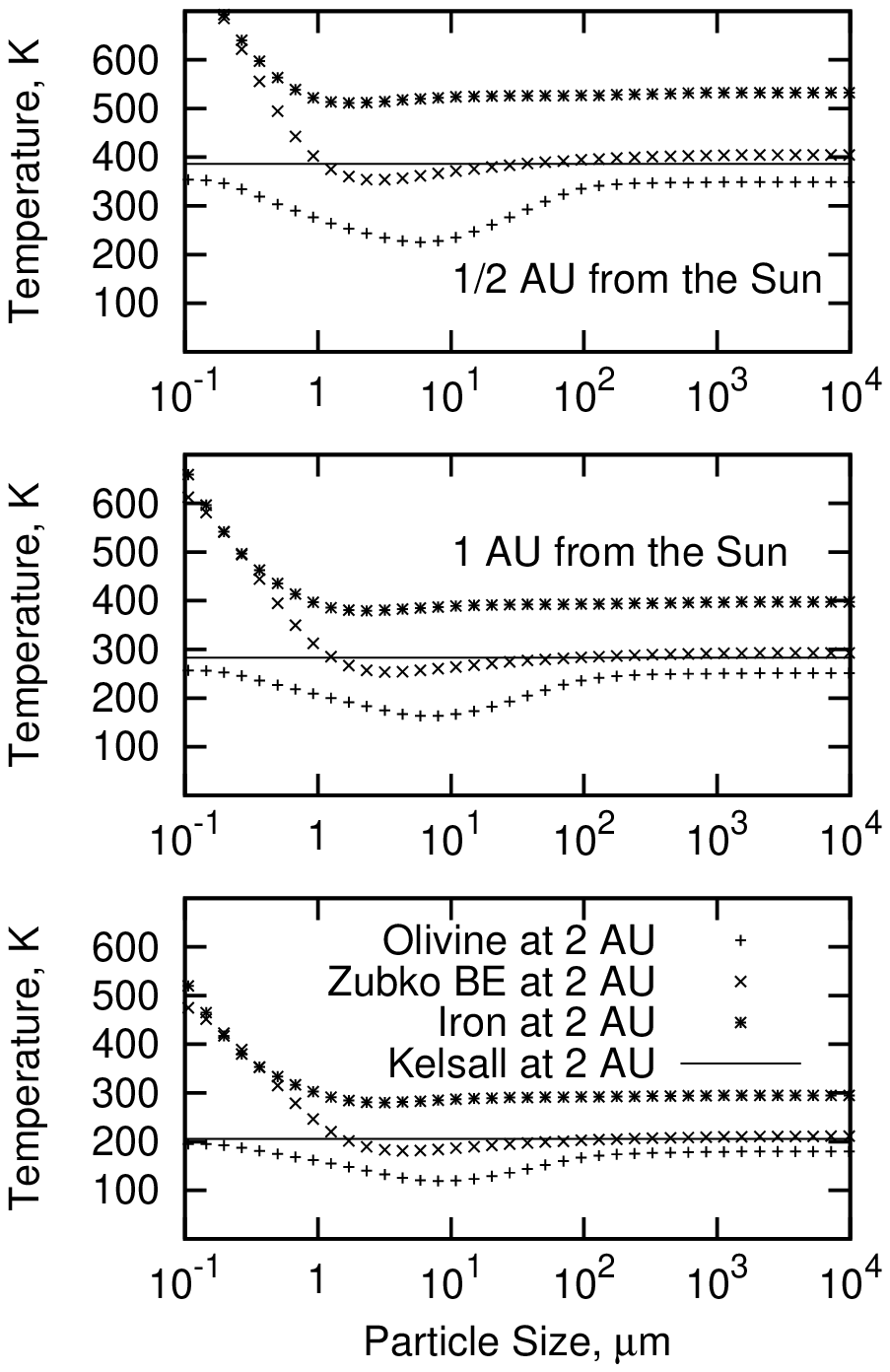}
\end{center}
\caption{Equilibrium temperatures of homogeneous spherical dust particles composed of different substances.\label{Equitemp-vs-size}}
\end{figure}

Amorphous carbonaceous particles (labeled ``Zubko BE'' in our plots) from $\sim10\;\mu$m in size are in the best
agreement with the Kelsall model temperatures, among the real species considered here. This is a good indication
for the Kelsall model, since most interplanetary meteoroids are indeed expected to be composed of
carbonaceous material~\citep{Reach-1988,Dikarev-et-al-2009}. The temperatures of the silicaceous particles
are somewhat lower, as olivine is rather transparent in the visual light that it absorbs, remaining opaque
in the infrared light that it emits, but this difference is quite negligible starting from particle radii of~$100\;\mu$m,
especially at the heliocentric distances beyond 1~AU (Fig.~\ref{Equitemp-vs-size}).

In subsequent sections, however, we assume a single temperature for all dust particles of each substance, regardless
of size. Our results are therefore less precise for dust particles smaller than $\sim100\;\mu$m in radii.
Fortunately for achievement of this paper's goal, the microwave emission from large particles
is more important to predict accurately, since small particles are not efficient emitters at
that long wavelengths.

\section{The Microwave Thermal Emission from Dust}\label{The Microwave Thermal Emission from Dust}

We are ready now to make maps of the thermal emission from the Zodiacal dust for
the broadest range of wavelengths and observation points. Indeed, the surface
brightness of the sky due to the Zodiacal cloud,
measured in units of power per unit solid angle and unit wavelength,
observed from the location~$\vec r$ in the direction specified by a unit vector~$\hat{\vec p}$,
is given by
\begin{equation}
\int_0^\infty B_\lambda(T_\mathrm{D}(|\vec r + l\hat{\vec p})|) C_\mathrm{abs} (\lambda, \vec r + l\hat{\vec p}) \mathrm{d}l,
\end{equation}
where $C_\mathrm{abs} (\lambda,\vec r)$ is the absorption cross-section of dust per unit volume of space.
In the Kelsall model, $C_\mathrm{abs}$ is a sum of the products of the total cross-section area of dust
by the emissivity modification factors for the wavelength~$\lambda$, over all model components.
Meteoroid engineering models require an evaluation of a nested integral over the particle size~$s$:
\begin{equation}\label{Cabs}
C_\mathrm{abs} (\lambda,\vec r) = -\int \pi s^2 \; Q_\mathrm{abs}(\lambda,s) \;
{\frac{\partial N(s, \vec r)}{\partial s}} \; \mathrm{d}s,
\end{equation}
where $N(s,\vec r)$ is the number density of meteoroids with radii greater than~$s$ at the
location $\vec r$. The minus sign in Eq.~(\ref{Cabs}) is necessary since
$\partial N(s,\vec r) / \partial s \leq 0$.
Figure~\ref{Equitemp-vs-size} shows that the equilibrium temperature
$T_\mathrm{D}$ depends on particle size as well, however, this dependence is important
only for the small dust grains with $s<100\;\mu$m, which can be safely neglected
in the microwaves.

Note that no color correction factor is applied, so that a difference
of at least several percent is to be expected between our intensities and
Kelsall's Zodiacal Atlas for the wavelengths where a direct comparison
is possible.

\subsection{All-Sky Maps}

\begin{figure*}
\begin{center}
\includegraphics[width=0.9\textwidth]{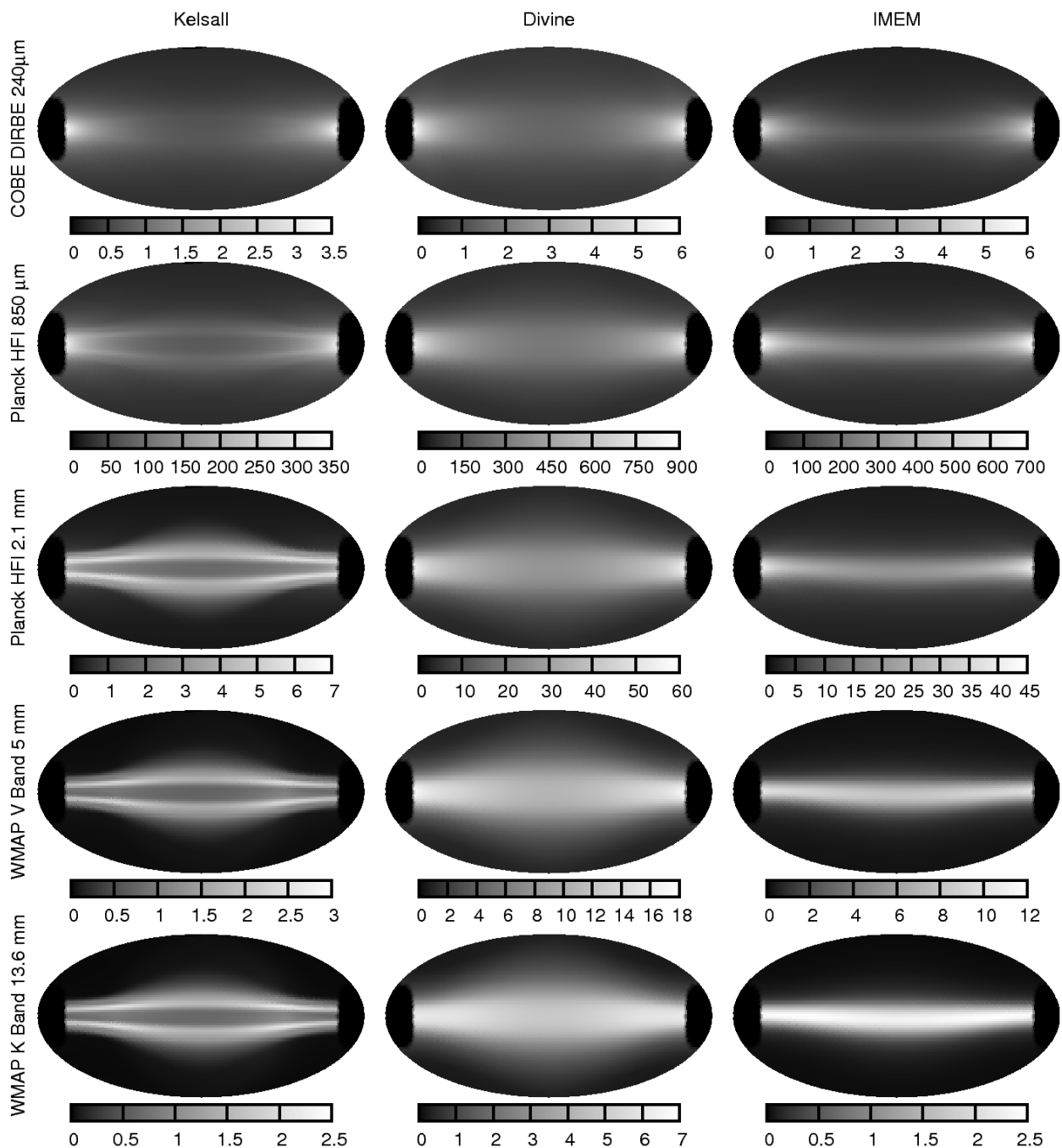}
\end{center}
\caption{All-sky maps of the thermal emission from the Zodiacal cloud, according to the Kelsall model and
two meteoroid engineering models~\citep{Divine-1993,Dikarev-et-al-2004EMP}
populated by the carbonaceous particles, as seen from Earth at the fall equinox time,
for selected wavebands of the {\it COBE}, {\it Planck} and {\it WMAP} observatories.
The reference frame is ecliptic, each map's center points to the antisolar direction
(i.e., the vernal equinox), a 30$^\circ$-wide band around the Sun is masked.
The grey scale of the upper row of maps is in MJy~sterad$^{-1}$,
the other rows are in $\mu$K of a temperature in excess of the CMB.
Each map has its own brightness scale.\label{All-sky maps C}}
\end{figure*}

All-sky maps of the thermal emission from the interplanetary dust are made
in Fig.~\ref{All-sky maps C} using the Kelsall model as well as meteoroid
engineering models assuming that their particles are carbonaceous. Five observation
wavebands of {\it COBE}, {\it Planck} and {\it WMAP} are selected to demonstrate
the transformation of the brightness distribution
from the far infrared light (240~$\mu$m) to microwaves (13.6~mm).
\begin{figure*}
\begin{center}
\includegraphics[width=0.9\textwidth]{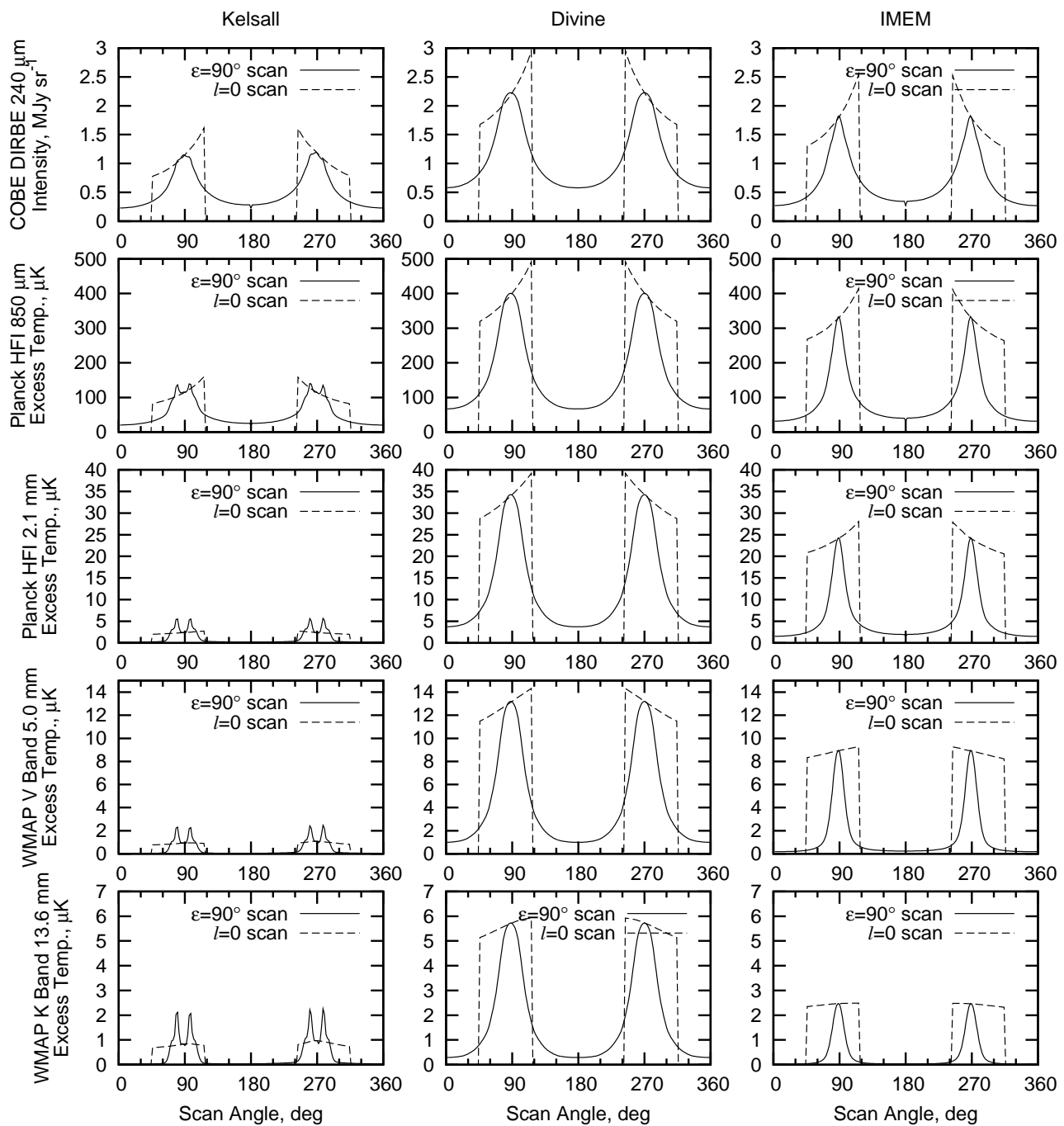}
\end{center}
\caption{Brightness profiles of the thermal emission from the Zodiacal cloud
along the great circles in the ecliptic plane (ecliptic latitude $l=0$) and perpendicular to
the ecliptic plane at solar elongation $\varepsilon=90^\circ$, according to the Kelsall model and
two meteoroid engineering models populated by the carbonaceous particles, as seen from Earth,
for selected wavebands of the {\it COBE}, {\it Planck} and {\it WMAP} observatories.
The excess temperatures are calculated w.r.t.\ the CMB.
\label{Profiles C}}
\end{figure*}

At short wavelengths, emission from fine dust grains is dominant,
and all models of the Zodiacal cloud show the maximum emission near the Sun.
The smooth background cloud is responsible for this brightness distribution
in the Kelsall model, the core population in the Divine model and the meteoroids
with masses~$m<10^{-5}$~g migrating toward the Sun under the Poynting-Robertson drag
in IMEM are also concentrated near the Sun.
As the wavelength grows, the fine dust grains fade out, but bigger meteoroids
remain bright. The components of the Kelsall model possessing spectra of macroscopic
particles, i.e.\ the asteroid dust bands, Divine's asteroidal population and
the meteoroids with masses~$m>10^{-5}$~g in IMEM are located mostly beyond the orbit of the Earth.
Instead of the brightness peak of the fine dust at small solar elongations,
the big meteoroids appear as a band, or bands, along the ecliptic, broadening in
the antisolar direction: the surface brightness of diffuse objects does not depend
on the distance from the observer, but their angular size increases due to their proximity
in the antisolar direction.

Figure~\ref{Profiles C} shows the brightness profiles built with the models from Fig.~\ref{All-sky maps C}
for Earth-bound observatories scanning the celestial sphere in two important great circles,
one in the ecliptic plane (zero ecliptic latitude) starting from the vernal equinox counterclockwise,
and another perpendicular to the ecliptic plane, at the solar elongation of~$90^\circ$,
from the ecliptic north pole toward the apex of Earth's motion about the Sun.
The scan in the ecliptic plane is only unmasked, however, when the target area on the sky
was visible with {\it WMAP} (solar elongations from~90$^\circ$ to~135$^\circ$) or {\it COBE} DIRBE
(solar elongations from~64$^\circ$ to~124$^\circ$). Note that both profiles match at scan angles
of 90$^\circ$ and 270$^\circ$ since the great circles are crossing there.

The Divine model and IMEM with the carbonaceous meteoroids appear to be substantially brighter than the Kelsall model.
Note that IMEM was fitted to the DIRBE data up to $\lambda=100\;\mu$m only,
using a composition of dust different from the triple used here,
and Divine used no infrared data at all. Thus no exact match with the Kelsall model is expected,
especially at the wavelengths $\lambda>100\;\mu$m for which \citet{Kelsall-et-al-1998} themselves
found the {\it COBE} DIRBE data insufficiently constraining the model (Sect.~\ref{sec:The Kelsall model}).
The difference is substantially lower (25--30\%) at shorter wavelengths (12--60~$\mu$m).
However, \citet{KenGanga-2013} inferred the absorption efficiencies of the Kelsall model
components for the {\it Planck} wavebands as well, and reported a lower emission
from the Zodiacal cloud as shown in Fig.~\ref{All-sky maps C} and~\ref{Profiles C}
on the plots for $\lambda=850\;\mu$m and~2.1~mm.

To pave the way for an explanation of the discrepancy, we remind that the interplanetary dust
is not entirely carbonaceous, and meteoroids of other chemical composition yield
lower infrared and microwave thermal emission. Besides, meteoroid engineering models
are fitted to many other data sets, typically with an accuracy substantially lower than
the models of the electromagnetic emission from dust. Finally, the fitting of the Kelsall
model to the {\it Planck} data by \citet{KenGanga-2013} resulted in confusingly negative
absorption efficiencies of dust sometimes, implying that the model may be incomplete,
with some model components used by the fitting procedure to compensate for excessive
emission from the other components. One can also see that the Kelsall model
is deficient in the ecliptic plane, with the asteroid dust bands allowed to shine
just a little above and below that plane only, and not e.g.\ in the ecliptic plane.
A new fit to the {\it Planck} data using different meteoroid
models would therefore be of great value, since these models may
be better in reproducing the relative brightness distribution
of the microwave emission from dust, but they also may currently
overestimate the number of big meteoroids in the interplanetary space,
so they also may be improved by such a fit.

Maps of the thermal emission of silicate particles are qualitatively similar with those
of the carbonaceous meteoroids and are not shown here.
However, the emission spectra of the zodiacal cloud for the silicate
particles are discussed in the next Section.

\begin{figure*}
\begin{center}
\includegraphics[width=0.9\textwidth]{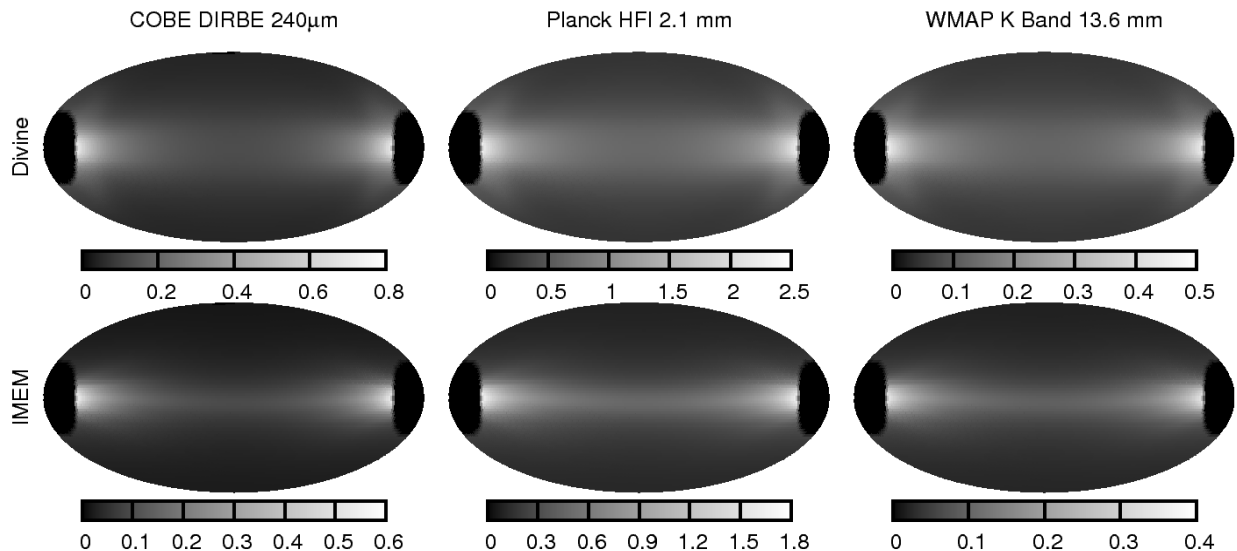}
\end{center}
\caption{All-sky maps of the thermal emission from the interplanetary dust predicted by two meteoroid
engineering models, assuming that their constituent particles are composed of iron.
The reference frame is as in Fig.~\ref{All-sky maps C}. The brightness is provided in units
of MJy~sterad$^{-1}$ for the {\it COBE} waveband centered at 240~$\mu$m,
and in units of $\mu$K for the {\it Planck} and {\it WMAP} wavebands.\label{All-sky maps Fe}}
\end{figure*}
We make maps of the thermal emission from the iron meteoroids since they are remarkably different
from the two materials discussed above~(Fig.~\ref{All-sky maps Fe}).
Fine dust grains bulking near the Sun remain the brightest source regardless of the wavelength of observation.
This stems from a very weak dependence of the absorption efficiency of iron particles on their size (within
the range considered in this paper, see Fig.~\ref{nkQabs}).

The difference in morphology of the maps for carbonaceous and iron particles suggests the way to discriminate
between different sorts of interplanetary meteoroids from the microwave observations, when their accuracy permits.


\begin{figure}
\begin{center}
\includegraphics[width=\columnwidth]{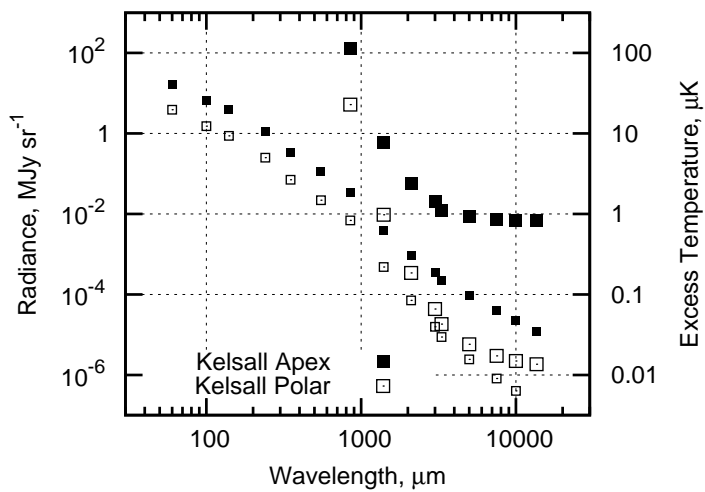}
\end{center}
\caption{Thermal emission spectra of the Zodiacal cloud predicted by the Kelsall model for the apex
of Earth's motion about the Sun and ecliptic pole. Large squares are the excess temperatures
w.r.t.\ the blackbody emission of the CMB (right scale), small squares are the absolute intensities
of emission (left scale).\label{Spectra Kelsall}}
\end{figure}

\subsection{Thermal Emission Spectra}\label{subsection Thermal Emission Spectra}

\begin{figure}
\begin{center}
\includegraphics[width=\columnwidth]{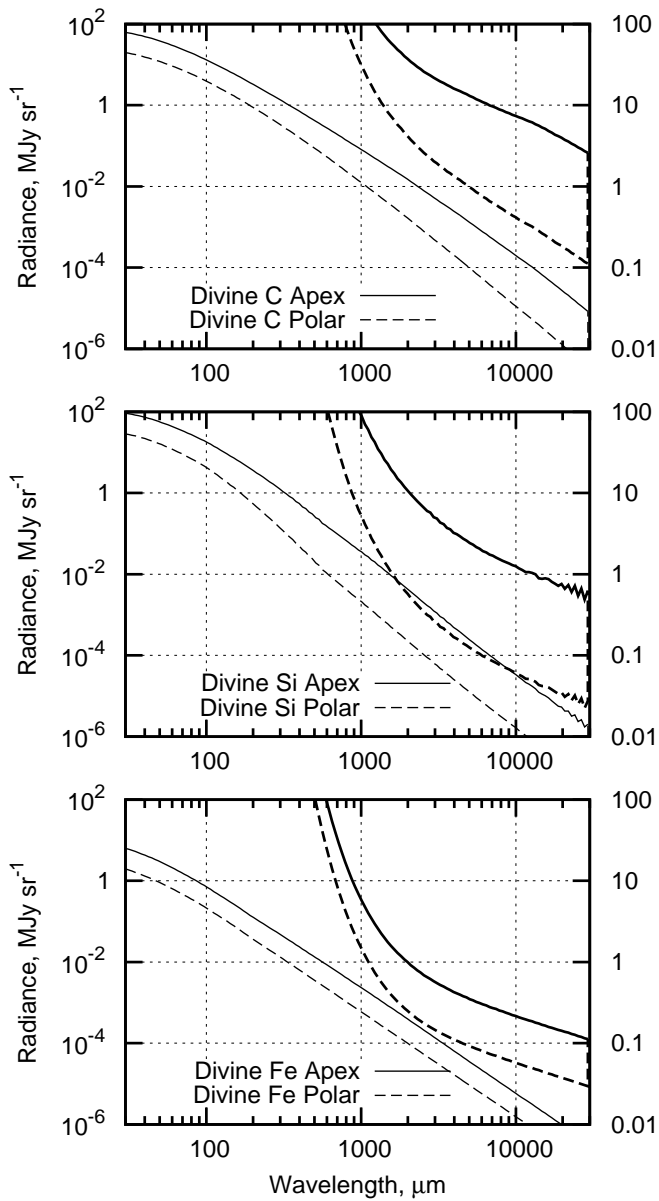}
\end{center}
\caption{Thermal emission spectra of the Zodiacal cloud predicted by the Divine model for the apex
of Earth's motion about the Sun and ecliptic pole, assuming that the dust particles
are composed of amorphous carbon (top), silicate (middle) and iron (bottom) material.
Thick curves are the excess temperatures w.r.t.\ the blackbody emission of the CMB (right scale),
dashed curves are the absolute intensities of emission (left scale).\label{Spectra Divine}}
\end{figure}

\begin{figure}
\begin{center}
\includegraphics[width=\columnwidth]{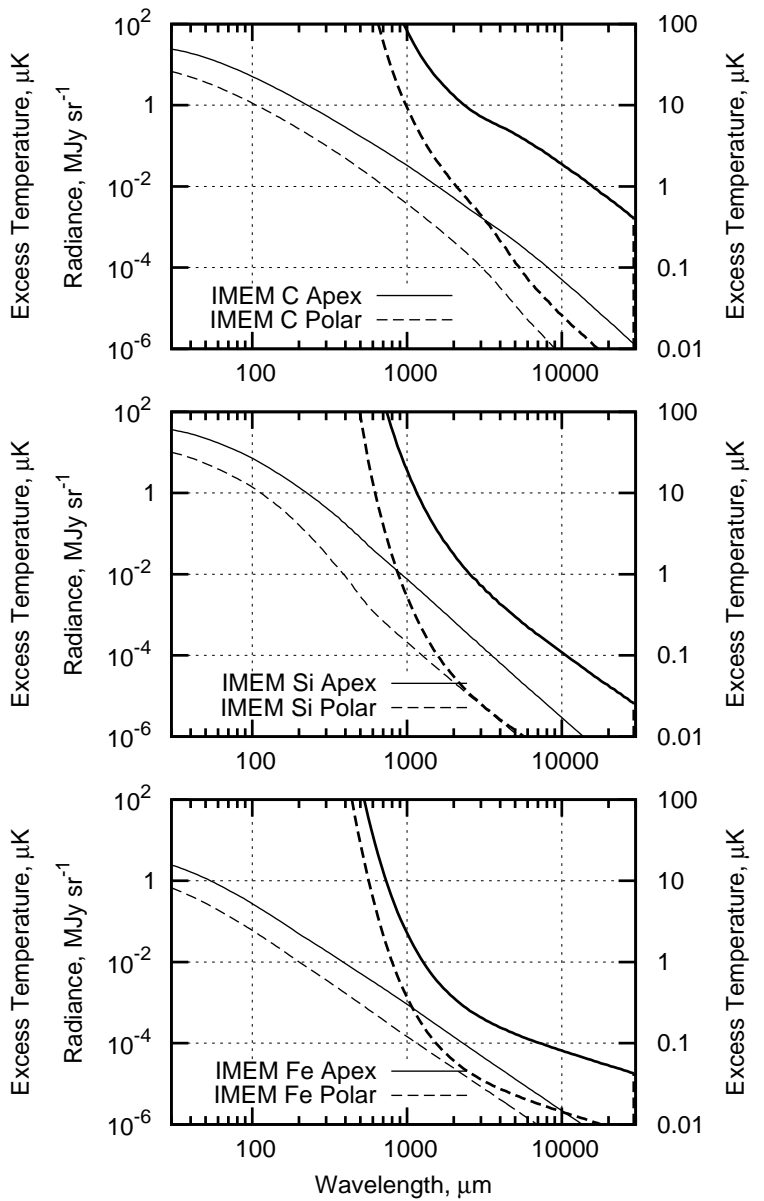}
\end{center}
\caption{Thermal emission spectra of the Zodiacal cloud predicted by IMEM for the apex
of Earth's motion about the Sun and ecliptic pole, assuming that the dust particles
are composed of amorphous carbon (top), silicate (middle) and iron (bottom) material.
Thick curves are the excess temperatures w.r.t.\ the blackbody emission of the CMB (right scale),
dashed curves are the absolute intensities of emission (left scale).\label{Spectra IMEM}}
\end{figure}

Figures~\ref{Spectra Kelsall}, \ref{Spectra Divine} and \ref{Spectra IMEM} show
the thermal emission spectra of the Zodiacal cloud predicted by Kelsall~et~al.,
pointed toward the apex of Earth's motion about the Sun and ecliptic pole.
The graphs are provided with two brightness scales: the absolute intensities
are more natural and easier to read in the infrared wavelengths, whereas the
excess temperatures are instantaneously comparable with other foreground
emission sources in the CMB studies. The excess temperatures~$\Delta T$ are defined
in accord with~\citet{Dikarev-et-al-2009} so that
$B_\lambda(T_\mathrm{CMB}+\Delta T) = B_\lambda(T_\mathrm{CMB}) + \epsilon$,
where $\epsilon$ is the intensity of emission in excess of the CMB radiation
at the temperature $T_\mathrm{CMB}=2.725$~K, with the blackbody intensity~$B_\lambda$
being calculated precisely with the Planck formula rather than in
the Rayleigh-Jeans approximation.

The Kelsall model is evaluated at the wavelengths of observations with
{\it COBE} DIRBE and {\it Planck} HFI, for which the absorption efficiencies
(``emissivity modification factors'') of the dust particles in the model components
have been determined, and using the extrapolations
described in Sect.~\ref{sec:Kelsall thermal} for the {\it WMAP} wavelengths. 
The model predicts rather low emission in the microwaves, e.g.\ below 1~$\mu$K in the apex and below 0.1~$\mu$K
in the ecliptic poles for $\lambda>3\;\mu$m. The emission in the apex is almost entirely due to the asteroid
dust bands presumably having flat absorption efficiencies at these wavelengths, while a continuing decrease
of the excess temperature with the wavelength increase for the polar line of sight for $\lambda>3$~mm
indicates that the smooth background cloud is contributing with $Q_\mathrm{abs}\propto\lambda^{-2}$.
According to the Kelsall model, the foreground emission due to the Zodiacal dust
could indeed be disregarded in the {\it WMAP} survey of the microwave sky with its
accuracy of 20~$\mu$K per $0.3^\circ$ square pixel over one year.

The Divine model in Fig.~\ref{Spectra Divine} turns out to be a lot brighter, however.
It predicts an excess temperature of up to $\sim20\;\mu$K in {\it WMAP}'s W~Band at $\lambda=3$~mm,
and $\sim5\;\mu$K in the K~Band at $\lambda=13.6$~mm, assuming that the meteoroids are all carbonaceous.
Silicate meteoroids would add much less than $10\;\mu$K in the entire {\it WMAP} wavelength range.
If the interplanetary dust particles were composed exclusively of iron, the Divine model would not
let them shine brighter than the Kelsall model.

Predictions made with IMEM for the same compositions as above are shown in Fig.~\ref{Spectra IMEM}.
IMEM is dimmer than the Divine model, it is still brighter than that by Kelsall.

%



We have also estimated the total mass and cross-section of meteoroids in the Divine model
and IMEM within 5.2~AU from the Sun, i.e.\ inside a sphere considered by \citet{Kelsall-et-al-1998}
as the outer boundary of the Zodiacal cloud. The Divine model contains $\sim2\cdot10^{20}$~g mass and
$6\cdot10^{-6}$~AU$^2$ cross-section area of meteoroids. Assuming a density of 2.5~g~cm$^{-3}$,
the total mass of meteoroids could be contained in a single spherical body
of the radius $\sim30$~km, the total cross-section area correspons to
a single sphere of the radius~$\sim2\cdot10^{5}$~km.
The last value of radius would be possessed by a planet
three times bigger than Jupiter.
IMEM has only $6\cdot10^{19}$~g mass and $6\cdot10^{-6}$~AU$^2$
cross-section area of meteoroids within 5.2~AU from the Sun.

\citet{Fixsen-Dwek-2002} assumed a certain size distribution of interplanetary meteoroids
and used the optical properties for silicate, graphite and amorphous carbonaceous
particles in order to reproduce the annually-averaged spectrum of the Zodiacal cloud
measured with the FIRAS instrument on board {\it COBE}. The smooth cloud of the Kelsall
model was used to describe the spatial number density distribution of meteoroids.
They found the total mass of meteoroids in the range 2--11$\cdot10^{18}$~g,
i.e.\ remarkably lower than in the Divine model and IMEM.


\section{Conclusion}\label{Conclusion}

We have made predictions of the microwave thermal emission from the Zodiacal dust cloud
in the wavelength range of {\it COBE} DIRBE, {\it WMAP} and {\it Planck} observations.
We used the \citet{Kelsall-et-al-1998} model extrapolated to the microwaves using
the optical properties of interplanetary dust inferred by \citet{KenGanga-2013}
and \citet{Fixsen-Dwek-2002}, and two engineering meteoroid models \citep{Divine-1993,Dikarev-et-al-2004EMP}
in combination with the Mie light-scattering theory to simulate the thermal emission of
silicate, carbonaceous and iron spherical dust particles.

We have found that the meteoroid engineering models depict the thermal emission substantially
brighter and distributed differently across the sky and wavelengths than the Kelsall model does,
due to the presence of large populations of macroscopic particles in the engineering meteroid models
that are only available in the asteroid dust bands of the Kelsall model. Both the Divine model and IMEM confirm an earlier
estimate of a $\sim10\;\mu$K thermal emission from interplanetary dust \citep{Dikarev-et-al-2009}
for the {\it WMAP} observations, provided that the dominant particle composition is carbonaceous.
At smaller solar elongations, interplanetary dust can be orders of magnitude brighter, naturally.

More detailed search for and account of interplanetary dust are therefore worthwhile in the CMB experiment results.
Maps of the microwave sky should also be scrutinized to constrain the number of macroscopic
particles in the engineering meteoroid models, as they are important to assess the impact hazard
for long-term manned missions with large spaceships to the Moon or Mars.

\begin{acknowledgements}
The authors thank Carlo Burigano, Craig Copi, Kenneth Ganga, Dragan Huterer, Pavel Naselsky and Glenn Starkman for insightful discussions.
This work was supported by the German Research Foundation (Deutsche Forschungsgemeinschaft), grant reference SCHW~1344/3~--~1,
and via the Research Training Group 1620 `Models of Gravity'. 
\end{acknowledgements}

\bibliography{icarus,dikarev}
\bibliographystyle{aa}
\end{document}